# Lindenmayer graph languages, first-order theories and expanders

Teodor Knapik[1]

[1]ISEA, Université de la Nouvelle Calédonie,

May 25, 2024

**Abstract**

Combinatorial generation of expander families and Lindenmayer-style development models are both parallel in nature. Both can be handled within proposed parallel graph grammar formalism. Their first-order properties can then be checked by encompassing the generated graph language into an appropriate automatic structure.

# 1 Introduction

Deeply concerned about the ongoing destructive processes that threaten the mankind and the nature, we advocate for an increased effort toward applications of relevant theories beyond computer science and mathematics. Formal languages, graphs and automata cover several themes in line with the current emergency. Lindenmayer systems, also known as L-systems, and expanders show promising potential for such applications as generative mechanisms for modelling structures of interest in materials science. Their connections to logic in computer science include the following question: "Which properties of expander families can be algorithmically checked?"

## Our contribution

To answer the question above, we introduce 0L graph grammars[1] and demonstrate how to implement the most important constructs for the combinatorial generation of expander families within our formalism. To remain credible, we design it to capture grids, meshes, and similar multidimensional structures that are essential in physics, chemistry, and biology. This application-oriented requirement excludes the possibility of checking monadic second-order definable properties. We only consider here those definable at the first order, nevertheless retaining the quantitative logics for a more in-depth future investigation.

We identify three features of 0L graph grammars that shape the border of the decidable: edge-determinism, vertex-determinism and completeness. After defining the structure encompassing the graph language generated by a 0L graph grammar, we show that its word-automatic presentation can be constructed in quadratic time. Thanks to our another construction that makes complete any 0L graph grammar, the completeness is not a restriction at this stage.

[1]In the classification of L-systems "0" stands for "context-free" and "L" for Lindenmayer.

Given a first-order sentence in graphs signature [7] and a 0L graph grammar, we ask if the corresponding 0L graph language includes a graph satisfying the sentence. By checking the associated automatic structure for a an adequate translation of the input sentence, we have an answer in the case of deterministic 0L graph grammars. On the other side of the decidability border, we reduce the acceptance of the empty word by a deterministic Turing machine to the latter question. The reduction yields a 0L graph grammar that is edge-deterministic and possesses a unique derivation. With a few satellite undecidability results, we can conclude that the challenge lies in identifying individual graphs within the structure encompassing the entire 0L graph language.

## Significance

Introduced in 1968 by Aristid Lindenmayer [25] and later generalised as graph grammars [20], L-systems remain an active research area [11]. Their extensions have been widely used for modelling the growth of living organisms or plant structure [33, 27], often using dedicated programming languages [31]. However, their applications go beyond life sciences and affect areas as varied as image synthesis [35, 33], multi-agent simulation [28] or engineering [3].

Later named by Pinsker [32] but introduced by Kolmogorov and Barzdin as early as 1967, expanders have their origins in Kolmogorov's interest in the structure of axonal connections in the brain [2]. Applications of expanders span many areas such as communication networks, error-correcting codes, pseudorandomnes, cryptography, computational complexity including the PCP theorem [8], derandomisation, computational group theory or number theory to name a few (see [19] and [26] for more details).

To our knowledge, parallel graph grammars and expanders[2] have been absent from the field of logic. This paper establishes their link with finite automata and finite [15, 24] as well as metafinite [14, 4] model theory.

## Related work

Parallel graph rewriting has a rich literature (see [10]) from pioneering work of [9] to more recent developments [37, 36] but all these approaches are context-sensitive. With an undecidable emptiness problem, an algorithmic checking of their properties, which is our main concern, seems very unlikely. Loosely related to that stream of work, our contribution can be seen as a simplification of formalisms of [20] and [21]. Graphs are considered there up to isomorphism which makes difficult referencing vertices or edges, requires long definitions and quite complicated proofs. With its first order logic-based core, our work extends both approaches. Additionally, it allows for multiple edges and loops. Several results, extending those of [20], are a by-product of the (un)decidability of model or language checking problems for 0L graph languages. Drawing our inspiration in the rich theory of monadic second order logic of graphs [7], we aim to stimulate an exploration of the connections between parallel graph languages and logics.

## Organisation of this paper

After some preliminaries, in Sect. 3 we review the basics of automatic structures. In Sect. 4 we introduce 0L graph grammars and state their typical language-theoretic decidability results. In

[2]except computation theory

Sect. 5 we review the most prominent combinatorial constructions of expanders and show how the corresponding deterministic or vertex-deterministic 0L graph grammars can be constructed. Logic related decidability results are stated in Sect. 6 together with the discussion about the central problem of this paper: the language checking problem. Most of our positive results follow from the construction of Sect. 7, namely the embedding of any complete 0L graph grammar into an automatic structure. In Sect. 8 we show how to simulate any deterministic Turing machine by an edge-deterministic 0L graph grammar. Short conclusion closes the main part of the paper and all missing proofs can be found in subsequent appendices.

# 2 Basic notations

Throughout the paper, placeholder "_" is often used to avoid specifying inessential parts in a notation like e.g. writing "let $x \xrightarrow{a} _$" in a more compact way than "let $x \xrightarrow{a} y$ for some $y$" while $y$ does not matter. As usual $[n]$ stands for $\{1, 2, \ldots, n\}$ with $[0] = \emptyset$. A tuple of length $n \in \mathbb{N}$ (also called an $n$-tuple) on a set $D$ is a map $\bar{t} : [n] \to D$. When $\bar{t}(i) = e_i$, for $i \in [n]$, a usual notation $\bar{t} = (e_1, \ldots, e_n)$ is often used. The 0-tuple is written $()$. Instead of $D^{[n]}$, the set of all $n$-tuples on $D$ is simply written $D^n$. A tuple $w$ of symbols from a finite *alphabet* $\Gamma$ is called a *word* and is written $w = a_1 a_2 \ldots a_n$ where $w(i) = a_i$, $a_i \in \Gamma$ and $w(\leq i) = a_1 a_2 \ldots a_i$, for $i \in [n]$. The length of $w$, written $|w|$, is $n$ and the word of length 0, called the *empty word*, is written $\varepsilon$ with $\Gamma^0 = \{\varepsilon\}$. The set of all words is written $\Gamma^*$, where $\Gamma^* := \bigcup_{n \in \mathbb{N}} \Gamma^n$.

Instead of being indexed by numbers in $[n]$, tuples may also be indexed by elements of a set $\mathscr{X}$. The family of subsets of $\mathscr{X}$ of size $k$ is written $\wp_k(\mathscr{X})$. Given $\mathcal{X} \in \wp_n(\mathscr{X})$, an $\mathcal{X}$-*tuple* $\bar{t}$ of a set $D$ is an element of $D^{\mathcal{X}}$. An $\mathcal{X}$-*relation* on $D$ is a set of $\mathcal{X}$-tuples. Operations on such relations are similar to those of relational algebra [22]. Let $\mathcal{X} \in \wp_n(\mathscr{X})$ and $\mathcal{Y} \in \wp_m(\mathscr{X})$. For a (possibly partial) function $\sigma : \mathcal{X} \rightharpoonup \mathcal{Y}$, define an *upturn* (*along* $\sigma$) of $\bar{t} \in D^{\mathcal{Y}}$ (resp. $\mathcal{R} \subseteq D^{\mathcal{Y}}$) as

$$\begin{array}{rcl} \sigma^{-1}(\bar{t}) & := & \{\bar{s} \in D^{\mathcal{X}} : \forall x \in \mathsf{dom}(\sigma)\ \bar{s}(x) = \bar{t}(\sigma(x))\}, \\ \big(\text{resp. } \sigma^{-1}(\mathcal{R}) & := & \bigcup_{\bar{t} \in \mathcal{R}} \sigma^{-1}(\bar{t})\big) \ . \end{array} \tag{2.1}$$

The upturn of $\mathcal{R}$ along an inclusion $\mathcal{X} \supseteq \mathcal{Y}$ is called *a cylindrification of* $\mathcal{R}$ *up to* $\mathcal{X}$ and is written $\mathcal{X}(\mathcal{R})$. The upturn along an inclusion $\mathcal{X} \subseteq \mathcal{Y}$ (resp. $\mathcal{Y} \setminus \mathcal{X} \subseteq \mathcal{Y}$) is a projection of $\mathcal{R}$ on (resp. out of) $\mathcal{X}$ and is written $\pi_{\mathcal{X}}(\mathcal{R})$ (resp. $\pi_{\exists \mathcal{X}}(\mathcal{R})$).

Let $\square \notin \Gamma$ be an additional symbol called *blank* and set $\Gamma_\square := \Gamma \cup \{\square\}$. A *convolution* $\otimes \overline{w}$ of an $\mathcal{X}$-tuple of words $\overline{w} \in (\Gamma^*)^{\mathcal{X}}$ is a word $u : [l] \to \Gamma_\square^{\mathcal{X}}$, with $l = \max_{x \in \mathcal{X}} |\overline{w}(x)|$, on the alphabet $\Gamma_\square^{\mathcal{X}}$ of $\mathcal{X}$-tuples of letters of $\Gamma_\square$, with $\otimes \overline{w} := u$ defined by

$$u(j)(x) = \begin{cases} \overline{w}(x)(j), & \text{if } 1 \leq j \leq |\overline{w}(x)|, \\ \square, & \text{if } |\overline{w}(x)| < j \leq l. \end{cases}$$

A *hole* in word $t \in (\Gamma_\square^{\mathcal{X}})^*$ is a position $i$ such that $t(i)(x) = \square$ for some component $x \in \mathcal{X}$ and there exists $j > i$ such that $t(j)(x) \neq \square$. A word is *hole-free* if it has no hole and a language on $(\Gamma_\square^{\mathcal{X}})^*$ is *hole-free* if it has only hole-free words. Note that the result of a convolution is always *hole-free*. Operation "$\otimes$" is extended to $\mathcal{X}$-relations: $\bigotimes R := \{\otimes \overline{w} : \overline{w} \in R\}$, for every $R \subseteq (\Gamma^*)^{\mathcal{X}}$. The inverse of the convolution $\otimes^{-1}$ is defined for hole-free words over $\Gamma_\square^{\mathcal{X}}$ in the usual way.

A *(finite)* $\mathcal{X}$-*automaton* [18, 17, 16] on $\Gamma$ is a (finite word) automaton on $\Gamma_\square^{\mathcal{X}}$. An $\mathcal{X}$-automaton $\mathscr{A}$ is *hole-free* if its language $\mathscr{L}(\mathscr{A})$ is hole-free. As the set of hole-free words on $\Gamma_\square^{\mathcal{X}}$ is a regular language, any $\mathcal{X}$-automaton may be restricted in such a way that it accepts only hole-free words. In the sequel, all $\mathcal{X}$-automata are assumed to be hole-free. Therefore, every (hole-free) $\mathcal{X}$-automaton $\mathscr{A}$ induces the relation $[\![\mathscr{A}]\!] \subseteq (\Gamma^*)^{\mathcal{X}}$ defined by $[\![\mathscr{A}]\!] := \otimes^{-1}(\mathscr{L}(\mathscr{A}))$. When $\mathcal{X}$ is clear from the context or irrelevant, we speak of tuple automata or, more specifically, of pair (when $|\mathcal{X}| = 2$) or triple (when $|\mathcal{X}| = 3$) automata. The arity $\mathsf{ar}(\mathscr{A})$ of an $\mathcal{X}$-automaton is the arity of the induced relation: $\mathsf{ar}(\mathscr{A}) := \mathsf{ar}[\![\mathscr{A}]\!]$. We assume that the reader is familiar with standard constructions on finite (word) automata notably Boolean operations and concatenation and we only give one construction specific to tuple automata. For $\mathcal{X} \in \wp_n(\mathscr{X})$ and $\mathcal{Y} \in \wp_m(\mathscr{X})$ let $\sigma\colon \mathcal{X} \rightharpoonup \mathcal{Y}$ be a (possibly partial) function and let $\mathscr{A}$ be an $\mathcal{Y}$-automaton, $\mathscr{A} = (\mathbb{Q}, \Delta, \mathbb{I}, \mathbb{F})$. An *upturn of $\mathscr{A}$ along $\sigma$* is an $\mathcal{X}$-automaton $\sigma^{-1}(\mathscr{A}) = (\mathbb{Q}, \sigma^{-1}(\Delta), \mathbb{I}, \mathbb{F}_{\square})$, where $\mathbb{F}_{\square}$ and $\sigma^{-1}(\Delta)$ are defined by:

$$\mathbb{F}_{\square} := \{q \in \mathbb{Q} \ : \ \exists f \in \mathbb{F}\ q \overset{(\square^{\mathcal{X}})^*}{\underset{\mathscr{A}}{\dashrightarrow}} f\},$$
$$\sigma^{-1}(\Delta) := \left\{p \overset{\bar{a}}{\to} q \ : \ \exists \bar{b}\ \left(p \overset{\bar{b}}{\to} q \in \Delta \wedge \bar{a} \in \sigma^{-1}(\bar{b})\right)\right\} .$$

Specific variants of the upturn for $\mathcal{Y}$-automata are written as for $\mathcal{Y}$-relations: $\mathcal{X}(\mathscr{A})$ is a cylindrification up to $\mathcal{X}$, $\pi_{\mathcal{X}}(\mathscr{A})$ (resp. $\pi_{\exists\mathcal{X}}(\mathscr{A})$) is a projection on (resp. out of) $\mathcal{X}$.

**Lemma 2.1.** *For every $\mathcal{Y}$-automaton and every (possibly partial) function $\sigma\colon \mathcal{X} \rightharpoonup \mathcal{Y}$, one has $[\![\sigma^{-1}(\mathscr{A})]\!] = \sigma^{-1}([\![\mathscr{A}]\!])$.*

# 3 Logic and structures

A *relational signature* is a set $\Pi$ of predicate symbols, each symbol $\varrho \in \Pi$ possessing its arity $\mathsf{ar}(\varrho) \in \mathbb{N}$. A $\Pi$-*relational structure* on a set $D$ is a map $\mathfrak{D}\colon \Pi \to \bigcup_{i \in \mathsf{ar}(\Pi)} \wp(D^i)$ such that every predicate symbol $\varrho \in \Pi$ is mapped into a relation $\mathfrak{D}(\varrho) \subseteq D^{\mathsf{ar}(\varrho)}$.

*First-order formulae on* $\Pi$ are built, from atomic formulae with predicate symbols of $\Pi$, using variables in $\mathscr{X}$, connectives $\neg, \wedge, \vee, \supset$ and quantifiers $\forall, \exists$. Without loss of generality, each quantified variable is assumed occurring free in the subformula. The set they form is written $\mathrm{FO}(\Pi)$. We allow grouping of any sequence of universal or existential quantifiers $Q\,x_1 \ldots, Q\,x_k$ with $Q \in \{\forall, \exists\}$ into *quantifier block* $Q\mathcal{X}$ where $\mathcal{X} = \{x_1 \ldots x_k\}$. Every atomic formula is considered as being of the form $\varrho(\sigma)$ where $\sigma\colon [\mathsf{ar}(\varrho)] \to \mathcal{X}$ and $\mathcal{X} \in \wp_m(\mathscr{X})$ with $m \leq \mathsf{ar}(\varrho)$. In other words, a tuple of variables of an atomic formula is seen as an explicit mapping. For instance, assuming $x, y \in \mathscr{X}$, the traditional writing $r(y, x, y)$ where $r \in \Pi$ is of arity 3 is seen as $r(\sigma)$ where $\sigma\colon [3] \to \{x, y\}$ with $\sigma(1) = y$, $\sigma(2) = x$ and $\sigma(3) = y$. When $\mathcal{X}$ is the set of free variables of a formula $\varphi$, the latter is often written $\varphi(\mathcal{X})$. As every $\mathrm{FO}(\Pi)$-formula can be transformed into an equivalent one without $\wedge, \supset$ nor $\forall$, we often give definitions or constructions for $\mathrm{FO}(\Pi)$-formulae using no other connectives than $\neg, \vee$ and no other quantifiers than $\exists$.

Every formula $\varphi(\mathcal{X}) \in \mathrm{FO}(\Pi)$ induces on a structure $\mathfrak{D}\colon \Pi \to \bigcup_{i \in \mathsf{ar}(\Pi)} \wp(D^i)$ an $\mathcal{X}$-relation $\mathfrak{D}(\varphi(\mathcal{X})) \subseteq D^{\mathcal{X}}$ defined inductively as follows

$$\begin{aligned}
\mathfrak{D}(\varrho(\sigma)) &:= \sigma^{-1} \circ \mathfrak{D}(\varrho), \qquad \text{where } \varrho \in \Pi,\ \sigma\colon [\mathsf{ar}(\varrho)] \to \mathcal{X} \text{ and } \sigma^{-1} \text{ is given by (2.1)},\\
\mathfrak{D}(\neg\varphi(\mathcal{X})) &:= D^{\mathcal{X}} \smallsetminus \mathfrak{D}(\varphi(\mathcal{X}))\\
\mathfrak{D}(\exists \mathcal{Z}\, \varphi(\mathcal{X})) &:= \pi_{\exists\mathcal{Z}}(\mathfrak{D}(\varphi(\mathcal{X})))\\
\mathfrak{D}(\varphi(\mathcal{X}) \vee \psi(\mathcal{Y})) &:= (\mathcal{X} \cup \mathcal{Y})\big(\mathfrak{D}(\varphi(\mathcal{X}))\big) \cup (\mathcal{X} \cup \mathcal{Y})\big(\mathfrak{D}(\psi(\mathcal{Y}))\big)
\end{aligned}$$

By an analogy between a structure $\mathfrak{D}$ and a database, a formula $\varphi(\mathcal{X})$ may be considered as a *(first-order) query* with corresponding relation $\mathfrak{D}(\varphi(\mathcal{X}))$ induced on $\mathfrak{D}$ being the result of the query.

## Automatic structures

An *automatic presentation* of a relational structure $\mathfrak{D}\colon \Pi \to \bigcup_{i \in \mathsf{ar}(\Pi)} \wp(D^i)$ is given by a tuple of automata on $\Gamma$, $\mathscr{A}(\mathfrak{D}) = \big(\mathscr{A}_\delta, \mathscr{A}_{\mathrm{eq}}, (\mathscr{A}(\varrho))_{\varrho \in \Pi}\big)$ where $\mathscr{A}_\delta$ is a usual finite automaton, $\mathscr{A}_{\mathrm{eq}}$ is a pair automaton, and, $\mathsf{ar}(\mathscr{A}(\varrho)) = \mathsf{ar}(\varrho)$, for each $\varrho \in \Pi$, together with a surjective mapping $g\colon \mathscr{L}(\mathscr{A}_\delta) \to D$ such that

- $(u, v) \in [\![\mathscr{A}_{\mathrm{eq}}]\!]$ if, and only if $g(u) = g(v)$, for all $u, v \in \mathscr{L}(\mathscr{A}_\delta)$,
- $(u_1, \ldots, u_{\mathsf{ar}(\varrho)}) \in [\![\mathscr{A}(\varrho)]\!]$ if, and only if $\big(g(u_1), \ldots, g(u_{\mathsf{ar}(\varrho)})\big) \in \mathfrak{D}(\varrho)$, for all $u_1, \ldots, u_{\mathsf{ar}(\varrho)} \in \mathscr{L}(\mathscr{A}_\delta)$.

$\mathscr{A}(\mathfrak{D})$ is also said to be *an automatic presentation of* $\mathfrak{D}$ *via* $g$. Such a presentation is said *injective* when $g$ is so. Then $\mathscr{A}_{\mathrm{eq}} = \mathscr{A}(=)$ is an automaton realising the identity on $\mathscr{L}(\mathscr{A}_\delta)$. A structure possessing an automatic presentation is called an *automatic structure*. Using an appropriate intersection of automata, every $\mathscr{A}(\varrho)$ can be arranged so that $[\![\mathscr{A}(\varrho)]\!] \subseteq [\![\mathscr{A}_\delta]\!]^{\mathsf{ar}(\varrho)}$. This is assumed for the sequel. Automatic structures have been introduced in 1976 by Hodgson [18, 17] and independently by Koussainov and Nerode in 1994 [23] (see also survey [13]).

### Construction of query automaton

Perhaps the most prominent property of automatic structures is that the result of every first-order query is encoded by a regular language. In other words, every relation that is FO-definable on an automatic structure may be encoded as a relation induced by a tuple automaton.

Let $\mathscr{A}(\mathfrak{D}) = \big(\mathscr{A}_\delta, \mathscr{A}_{\mathrm{eq}}, (\mathscr{A}(\varrho))_{\varrho\in\Pi}\big)$ be an automatic presentation of a relational structure $\mathfrak{D}\colon \Pi \to \bigcup_{i\in\mathsf{ar}(\Pi)} \wp(D^i)$. To every formula $\varphi(\mathcal{X}) \in \mathrm{FO}(\Pi)$, an automaton $\mathscr{H}(\varphi(\mathcal{X}))$, called a *query automaton of* $\varphi(\mathcal{X})$, is associated inductively as follows

$$\begin{aligned}
\mathscr{H}(\varrho(\sigma)) &:= \sigma^{-1}(\mathscr{A}(\varrho)) && \text{where } \varrho \in \Pi \text{ and } \sigma\colon [\mathsf{ar}(\varrho)] \to \mathcal{X}\\
\mathscr{H}(\neg\varphi(\mathcal{X})) &:= \mathscr{A}_\delta^{\mathcal{X}} \smallsetminus \mathscr{H}(\varphi(\mathcal{X})),\\
\mathscr{H}(\varphi(\mathcal{X}) \vee \psi(\mathcal{Y})) &:= (\mathcal{X}\cup\mathcal{Y})(\mathscr{H}(\varphi(\mathcal{X}))) \cup (\mathcal{X}\cup\mathcal{Y})(\mathscr{H}(\psi(\mathcal{Y}))),\\
\mathscr{H}(\exists \mathcal{Z}\, \varphi(\mathcal{X})) &:= \pi_{\exists\mathcal{Z}}(\mathscr{H}(\varphi(\mathcal{X})))
\end{aligned}$$

where $\mathscr{A}_\delta^{\mathcal{X}}$ is an $\mathcal{X}$-automaton such that $[\![\mathscr{A}_\delta^{\mathcal{X}}]\!] = [\![\mathscr{A}_\delta]\!]^{\mathcal{X}}$, viz., $\mathcal{X}$-indexed free product of $|\mathcal{X}|$ copies of $\mathscr{A}_\delta$. The fundamental property of $\mathscr{H}$ is that $g([\![\mathscr{H}(\varphi(\mathcal{X}))]\!]) = \mathfrak{D}(\varphi(\mathcal{X}))$, for every $\varphi(\mathcal{X}) \in \mathrm{FO}(\Pi)$. In particular, for a sentence $\psi \in \mathrm{FO}(\Pi)$, one has $\mathfrak{D} \models \psi \Leftrightarrow \mathscr{L}(\mathscr{H}(\psi)) \neq \varnothing$.[3] This leads to the well-known result that the FO-model checking (resp. query evaluation) for automatic structures is decidable (resp. computable), even in the case when $\mathrm{FO}(\Pi)$ is extended with counting quantifiers saying that "there are $k$ mod $m$ many elements" or "there are infinitely many elements". Indeed, the above inductive construction of the query automaton can be extended with two corresponding cases [4]. The results of our paper remain valid in that extension.

### Remarks on complexity

The construction of query automaton, also known as query evaluation problem, has a reputation of non-elementary complexity [13]. However, for quantifier-free formula, this construction can be done in time $\mathcal{O}(m^n)$ where $m$ is the number of transitions of the largest automaton of $\mathscr{A}(\mathfrak{D})$ and $n$ is the number of atoms and their negations in the input formula [4]. In the general case, input formula can be put in prenex form in linear time. Then the time complexity is an exponential tower of height $k$ with $m^n$ on top of it, where $k$ is the number of quantifier alternations. Note that this worst-case complexity should not be considered prohibitive for real life applications. Properties that need four quantifier alternations or more are very uncommon. Also uncommon is the exponential blow-up of the determinisation. As is observed in [12], it occurs only in cases similar to the usual example.

[3] Note that, for a sentence $\psi \in \mathrm{FO}(\Pi)$, $\mathscr{H}(\psi)$ is an automaton over single letter alphabet $\{()\}$.

# 4 Lindenmayer graph languages

The grammatical formalism developed in this section subsumes those of graph grammars discussed in [21]. A 0L graph grammar introduced in the sequel generates (concrete) graphs as incidence structures of the form $(V, E, \ell, \daleth)$ where

- $V \subseteq \Theta^+$ are the vertices given as words over an alphabet $\Theta$ of *vertex names*,
- $E \subseteq \Theta^*\Omega^+$ are the edges given as words over an alphabet $\Omega$ of *edge names*, $\Theta \cap \Omega = \varnothing$, possibly with a prefix in $\Theta^*$,
- $\ell : V \to \Gamma$ is the labelling of vertices and
- $\daleth \subseteq \Sigma \times V \times E \times V$ is the incidence relation given as a set of *labelled incidences* of the form $a(u, w, v)$ with $u, v \in V$, $w \in E$ and $a \in \Sigma$, such that $\{a(u, w, v) \mapsto w \; : \; a(u, w, v) \in \daleth\}$ is a bijection between $\daleth$ and $E$.

The set of (concrete) graphs over vertex and edge alphabets $\Theta$ and $\Omega$ with $\Gamma$-labelled vertices and $\Sigma$-labelled edges is written $\mathscr{G}(\Gamma, \Sigma, \Theta, \Omega)$. The set of those graphs of $\mathscr{G}(\Gamma, \Sigma, \Theta, \Omega)$ that have their edges and vertices all of length $l \in \mathbb{N}$ is written $\mathscr{G}_l(\Gamma, \Sigma, \Theta, \Omega)$. This formalism is general enough to handle unlabelled or undirected graphs too. Undirected incidences $a(\{u, v\}, w)$ (resp. with labels at both extremities $(\{(a, u), (b, v)\}, w))$ often drawn as $u \overset{c}{—} v$ (resp. $u \overset{a \quad b}{—} v$) are represented by two opposite directed incidences (resp. with a pair of labels): $a(u, w, v)$ and $a(v, w, u)$ (resp. $(a, b)(u, w, v)$ and $(b, a)(v, w, u)$).

A *0L graph grammar* over a (finite) set $\Gamma$ of vertex labels, a (finite) set $\Sigma$ of edge labels, a (finite) set $\Theta$ of vertex names and a (finite) set $\Omega$ of edge names is a triple $\mathcal{G} = (\mathsf{V}_{\mathcal{G}}, \mathsf{E}_{\mathcal{G}}, \mathsf{A}_{\mathcal{G}})$ where

- $\mathsf{A}_{\mathcal{G}}$ is the axiom of the grammar representing a graph with no edge and single vertex encoded by the empty word labelled $\mathsf{A}_{\mathcal{G}}$ which can but is not necessarily in $\Gamma$,
- $\mathsf{V}_{\mathcal{G}}$ is a finite set of *vertex productions* of the form $A \to G$, with $A \in \Gamma \cup \{\mathsf{A}_{\mathcal{G}}\}$ and $G \subseteq \mathscr{G}_1(\Gamma, \Sigma, \Theta, \Omega)$, such that the right hand sides of $\mathsf{V}_{\mathcal{G}}$ are pairwise disjoint, respectively to their vertex and edge sets,[4]
- $\mathsf{E}_{\mathcal{G}}$ is a finite set of *edge productions* of the form $(A_1 \to G_1) \xrightarrow{a} (A_2 \to G_2) \;\; \to \;\; \sqsupset$, with $\sqsupset$ using two disjoint sets of edge labels $\overrightarrow{\Sigma}$ and $\overleftarrow{\Sigma}$, each one in bijection with $\Sigma$, where

$$\begin{aligned} a &\in \Sigma, \\ A_1 \to G_1,\ A_2 \to G_2 &\in \mathsf{V}_{\mathcal{G}},\ \text{with } G_1 = (V_1, _, _, _) \text{ and } G_2 = (V_2, _, _, _), \\ \text{and } \sqsupset &\subseteq (\Sigma \times V_1 \times \Omega \times V_2) \cup (\Sigma \times V_2 \times \Omega \times V_1) \text{ are production's incidences} \end{aligned}$$

such that every edge name of $\Omega$ appears at most once in the right hand sides of grammar's productions $\mathsf{V}_{\mathcal{G}} \cup \mathsf{E}_{\mathcal{G}}$.[4] The set of such grammars is written $0\mathrm{L}(\Gamma, \Sigma, \Theta, \Omega)$. The edge label from $\overrightarrow{\Sigma}$ (resp. $\overleftarrow{\Sigma}$) corresponding to $a \in \Sigma$ is written $\overrightarrow{a}$ (resp. $\overleftarrow{a}$). For every production $P \in \mathsf{V}_{\mathcal{G}} \cup \mathsf{E}_{\mathcal{G}}$ let $\mathrm{lhs}(P)$ (resp. $\mathrm{rhs}(P)$) stands for its left (resp. right) hand side, viz., for vertex production $P_1 = A \to G$, $\mathrm{lhs}(P_1) = A$ (resp. $\mathrm{rhs}(P_1) = G$) and for edge production $P = P_1 \xrightarrow{a} P_2 \to \sqsupset$, $\mathrm{lhs}(P) = P_1 \xrightarrow{a} P_2$ (resp. $\mathrm{rhs}(P) = \sqsupset$). As every symbol $i \in \Theta \cup \Omega$ appears at most once in $\mathrm{rhs}(\mathsf{V}_{\mathcal{G}} \cup \mathsf{E}_{\mathcal{G}})$, we denote by $\mathrm{prd}(i)$ the production $P \in \mathsf{V}_{\mathcal{G}} \cup \mathsf{E}_{\mathcal{G}}$ where $i$ appears in $\mathrm{rhs}(P)$. When $\mathsf{A}_{\mathcal{G}} \notin \Gamma$, a vertex production with the axiom on its left hand side is called *an axiom production.* A vertex (resp. edge) production of the form $A \to (\varnothing, _, _, _)$ (resp. $P \xrightarrow{a} P' \to \varnothing$) is called *erasing.* A grammar with no erasing productions is called *non-erasing.*

Every 0L graph grammar $\mathcal{G} = (\mathsf{V}_{\mathcal{G}}, \mathsf{E}_{\mathcal{G}}, \mathsf{A}_{\mathcal{G}})$ can be seen as a graph with vertices $\mathsf{V}_{\mathcal{G}}$ labelled in $\Gamma$ and edges $\mathrm{lhs}(\mathsf{E}_{\mathcal{G}})$ labelled in $\Sigma$ together with an additional information in $\mathrm{rhs}(\mathsf{E}_{\mathcal{G}})$. Every vertex production $A{\to}H$ is an $A$-labelled vertex of graph $\mathcal{G}$. Every edge production $(A{\to}H) \xrightarrow{a} (A'{\to}H') \;\to\; \sqsupset$ defines an $a$-labelled edge $(A{\to}H) \xrightarrow{a} (A'{\to}H')$ of graph $\mathcal{G}$. With this view of $\mathcal{G}$, from a (concrete) graph $G = (V, E, \ell, \daleth) \;\in\; \mathscr{G}_l(\Gamma, \Sigma, \Theta, \Omega)$ with vertices and edges of length $l$, one

[4] This requirement is not necessarily respected in examples in order to keep them simple.

*derives* in a single step a (concrete) graph $G' = (V', E', \ell', \gimel') \in \mathscr{G}_{l+1}(\Gamma, \Sigma, \Theta, \Omega)$ upon selecting a graph homomorphism $h\colon G \to \mathcal{G}$ satisfying $h(v) = \ell(v) \to _$ for every vertex $v \in V$ and, for every edge $w \in E$ such that $a(u, w, v) \in \gimel$, satisfying $h(w) = h(u) \xrightarrow{a} h(v) \to _$. Resulting graph $G'$ is said to be *produced by $h$ from $G$*, written $G \underset{\mathcal{G}}{\overset{h}{\to}} G'$, or *derived (in a single step) from $G$*, written $G \underset{\mathcal{G}}{\to} G'$ and is obtained as follows:

$$\begin{array}{rcl}
\ell' & := & \bigcup_{v \in V} \{ vi \mapsto \ell_v(i) \;:\; h(v) = _ \to (V_v, _, \ell_v, _) \;\wedge\; i \in V_v \}, \\
V' & := & \bigcup_{A \in \Gamma} (\ell')^{-1}(A), \\
\gimel' & := & \{ b(ui, wj, vk) \;:\; (h(w) = h(u) \xrightarrow{a} h(v) \to \sqsupset \;\wedge\; \overrightarrow{b}(i,j,k) \in \sqsupset) \} \cup \\
 & & \{ b(vk, wj, ui) \;:\; (h(w) = h(u) \xrightarrow{a} h(v) \to \sqsupset \;\wedge\; \overleftarrow{b}(i,j,k) \in \sqsupset) \} \cup \\
 & & \{ b(ui, uj, uk) \;:\; h(u) = _ \to (_, _, _, \gimel_u) \;\wedge\; b(i,j,k) \in \gimel_u \}, \\
E' & := & \{ w' \;:\; b(_, w', _) \in \gimel' \} \;.
\end{array}$$

In the above definition of $\gimel'$, incidences and corresponding edges in the 1st or 2nd (resp. in the 3rd) member of the union are called *inherited* (resp. *innate*). Every $ui \in V' \cup E'$ *derives (in a single step) from $u \in V \cup E$ via $P$*, written $v \underset{\mathcal{G}}{\overset{P}{\to}} vi$, where $P \in \mathsf{V}_{\mathcal{G}} \cup \mathsf{E}_{\mathcal{G}}$ is a production such that $h(u) = P$. If $u_0 \underset{\mathcal{G}}{\overset{P_1}{\longrightarrow}} \cdots \underset{\mathcal{G}}{\overset{P_n}{\longrightarrow}} u_n$, we write $u_0 \underset{\mathcal{G}}{\overset{P_1 \ldots P_n}{\longrightarrow}} u_n$ and say that $u_n$ *derives from $u_0$ via (sequence of productions) $P_1 \ldots P_n$*. The transitive closure of the single-step derivation relation over graphs is the *derivation relation*, written "$\underset{\mathcal{G}}{\overset{+}{\to}}$", and the *language of $\mathcal{G}$*, viz., the *set of graphs derived* or *generated by $\mathcal{G}$* is defined as usual: $\mathscr{L}(\mathcal{G}) := \{ G \;:\; \mathsf{A}_{\mathcal{G}} \underset{\mathcal{G}}{\overset{+}{\to}} G \}$.

A slight generalisation of $\mathscr{L}(\mathcal{G})$, written $\mathscr{L}(\mathcal{G})_N$, arises when one distinguishes a subset of *non-terminal* labels $N \subseteq \Gamma \cup \Sigma$ not to appear in the final result: $\mathscr{L}(\mathcal{G})_N := \mathscr{L}(\mathcal{G}) \cap \mathscr{G}(\Gamma \setminus N, \Sigma \setminus N, \Theta, \Omega)$. A label in $\Gamma \cup \Sigma$ (resp. production of $\mathcal{G}$) is *accessible* in (resp. is *useful* for) $\mathcal{G}$, if it appears in (resp. is used in a derivation of) a graph of $\mathscr{L}(\mathcal{G})$. A graph grammar is *trimmed* if it has only useful productions.A 0L graph grammar is *edge-deterministic* (resp. *vertex-deterministic*) and called a *eD0L* (resp. *vD0L*) *graph grammar* if, for every triple $(A_1, a, A_2) \in \Gamma \times \Sigma \times \Gamma$ (resp. vertex label $A \in \Gamma$), it has at most one edge $(A_1 \to _) \xrightarrow{a} (A_2 \to _) \to _ \in \mathsf{E}_{\mathcal{G}}$ (resp. one vertex $A \to _ \in \mathsf{V}_{\mathcal{G}}$). A 0L graph grammar is *deterministic* and is called *D0L graph grammar* if it is both vD0L and eD0L. The subset of $0\mathrm{L}(\Gamma, \Sigma, \Theta, \Omega)$ that consists of D0L graph grammars is denoted by $\mathrm{D0L}(\Gamma, \Sigma, \Theta, \Omega)$. A 0L graph grammar is *unix* if for every $l \in \mathbb{N}$ it has at most one derivation of length $l$. Obviously, every D0L graph grammar is unix.

We say that $\mathcal{G}$ is *complete*, if it satisfies the following two conditions:

(1) for every edge label $a \in \Sigma$ and all vertices $A_1 \to H_1, A_2 \to H_2 \in \mathsf{V}_{\mathcal{G}}$ such that $\mathsf{A}_{\mathcal{G}} \notin \{A_1, A_2\}$ and an incidence $a(u, _, v)$ with $u$ labelled $A_1$ and $v$ labelled $A_2$ appear a right hand side of a production, there exists an edge $(A_1 \to H_1) \xrightarrow{a} (A_2 \to H_2) \to _ \in \mathsf{E}_{\mathcal{G}}$,

(2) for every isolated vertex with label $A \in \Gamma$ appearing in a right hand side of a production, there exists a vertex $A \to _ \in \mathsf{V}_{\mathcal{G}}$.

Derivations according to a complete 0L graph grammar are "easy", in the sense that every choice of vertex productions to apply on some graph $G$, for deriving in a single step a "child", can be completed with a compatible choice of edge productions. This is formalised in the following lemma.

**Lemma 4.1.** *Let $\mathcal{G} = (\mathsf{A}_{\mathcal{G}}, \mathsf{V}_{\mathcal{G}}, \mathsf{E}_{\mathcal{G}})$ be a complete graph grammar in $0\mathrm{L}(\Gamma, \Sigma, _, _)$ and let $G = (V, E, \ell, \gimel)$ be a graph in $\mathscr{L}(\mathcal{G})$. Every labelling preserving map $h\colon V \to \mathsf{V}_{\mathcal{G}}$ extends into a homomorphism from $G$ into $\mathcal{G}$.*

The following proposition asserts that the completeness can be bypassed with the help of one non-terminal vertex label and one non-terminal edge label.

**Proposition 4.2.** *For every 0L graph grammar* $\mathcal{G} = (\mathsf{V}_\mathcal{G}, _, _)$ *one can construct in time* $\mathcal{O}(|\mathsf{V}_\mathcal{G}|^2)$ *using additional non-terminal vertex label "¡" and non-terminal edge label "!" complete 0L graph grammar* $\mathcal{G}^{¡!}$ *such that* $\mathscr{L}(\mathcal{G}) = \mathscr{L}(\mathcal{G}^{¡!})_{\{¡,!\}}$. *If* $\mathcal{G}$ *is vD0L or eD0L, then* $\mathcal{G}^{¡!}$ *is so.*

Note that, in general, the construction does not preserve the property of being unix, even for unix eD0L graph grammars.

The following graphical formalism is used in subsequent examples. Every edge production $(A_1 \to H_1) \xrightarrow{a} (A_2 \to H_2) \to \sqsupset$ in $\mathsf{E}_\mathcal{G}$ is drawn separately using colours as an (abstract) edge $A_1 \xrightarrow{a} A_2$ on the left hand side and $H_1 \cup H_2$ augmented with $\sqsupset$ on the right hand side. Vertex productions $A_1 \to H_1$ and $A_2 \to H_2$ are implicit form colour codes and are not listed unless they cannot be inferred, notably when the axiom uses an extra label not in $\Gamma$. Incidences $\sqsupset$ are drawn as black edges. In addition, vertex or edge names from $\Theta \cup \Omega$ can be mentioned in parentheses. During a rewriting step of a graphical representation of a graph $(V, E, \ell, \rceil)$, every incidence in $\rceil$ together with corresponding vertex labels is matched to a left hand side of a graphical representation of an edge production so that the matching agrees on common vertices. In other words, for each pair of incident edges drawn as e.g. $A \xrightarrow{a} B \xrightarrow{b} C$, the vertex production used at $B$ must be common to both edge productions applied on $A \xrightarrow{a} B$ and $B \xrightarrow{b} C$ (and similarly for other orientations of edges like $A \xleftarrow{a} B \xrightarrow{b} C$ or $A \xrightarrow{a} B \xleftarrow{b} C$). This is a graphical counterpart of the homomorphism condition in the definition of derivation.

Note that arrows of $\overrightarrow{\Sigma}$ and $\overleftarrow{\Sigma}$ are not needed in a graphical representation. Moreover, even a formal statement of $\mathcal{G}$ can be simplified by removing arrows of $\overrightarrow{\Sigma}$ and $\overleftarrow{\Sigma}$ in every edge production, the left hand side of which is not a loop. If however it is a loop, like in e.g. $(A \to H) \xrightarrow{a} (A \to H) \to \sqsupset$, when homomorphism $h{:}\, G \to \mathcal{G}$ producing $G'$ from $G$ maps to it an edge between two distinct vertices of $G$, then $\overrightarrow{\Sigma}$ and $\overleftarrow{\Sigma}$ are necessary to avoid a confusion about the direction of edges created by $\sqsupset$ in $G'$ between the two copies of $H$. Now, in the corresponding graphical representation of $\mathcal{G}$ we draw an edge $A \xrightarrow{a} A$ at the left hand side and there is no ambiguity about the direction of edges added on the right hand side between the two copies $H$ and $H$. Nevertheless, in such a case, one needs to keep in mind that, in a graphical representation of the above edge production, $A \xrightarrow{a} A$ corresponds to a loop of $\mathcal{G}$ which can be matched by loop of $G$ over one of its vertices, say $v$. Then, in produced $G'$, there is single copy of $H$ corresponding to $v$ (instead of two copies of $H$, in case the matched edge of $G$ is not a loop) and edges created by $\sqsupset$ for that copy are loops.

Another difference between a formal statements and a graphical formalism concerns the meaning of "right hand side". For edge production $P = (A_1 \to H_1) \xrightarrow{a} (A_2 \to H_2) \to \sqsupset$, $\mathrm{rhs}(P)$ appears only as a portion of the right hand side of the graphical representation of $P$. Indeed, the latter includes drawings of $H_1$ and $H_2$ whereas the former is $\mathrm{rhs}(P) = \sqsupset$. The following example illustrates the differences between the graphical representation and the formal statement of a 0L graph grammar.

**Example 4.3**
*Consider* $\Gamma := \{A\}$, $\Sigma := \{a\}$, $\Theta := \{0,1,2,3,4\}$, $\Omega := \{\alpha,\beta,\gamma,\delta\}$ *and* $(\mathsf{Ax},\mathsf{V},\mathsf{E}) \in 0\mathrm{L}(\Gamma,\Sigma,\Theta,\Omega)$ *where*

$$\begin{aligned}
\mathsf{V} &:= \{\mathsf{Ax} \to G_0,\ A \to G_1,\ A \to G_2\},\\
\mathsf{E} &:= \big\{(A \to G_1) \xrightarrow{a} (A \to G_1) \to \sqsupset_1,\ \ (A \to G_2) \xrightarrow{a} (A \to G_2) \to \sqsupset_2\big\}
\end{aligned}$$

*with*

$$\begin{aligned}
G_0 &:= \big(\{0,1,2\},\{\alpha,\beta\},\{0 \mapsto A, 1 \mapsto A, 2 \mapsto A,\},\{a(0,\alpha,1), a(1,\beta,2)\}\big),\\
G_1 &:= \big(\{3\},\varnothing,\{3 \mapsto A,\},\varnothing\big), \qquad \sqsupset_1 := \{\overrightarrow{a}(3,\gamma,3)\},\\
G_2 &:= \big(\{4\},\varnothing,\{4 \mapsto A,\},\varnothing\big), \qquad \sqsupset_2 := \{\overleftarrow{a}(4,\delta,4)\}\ .
\end{aligned}$$

(Ax, V, E) *is represented graphically on the right. It consists of the axiom production and the two edge productions. Remaining vertex productions, namely $A \to A(3)$ and $A \to A(4)$, are omitted, as these can be inferred from the two edge productions.*

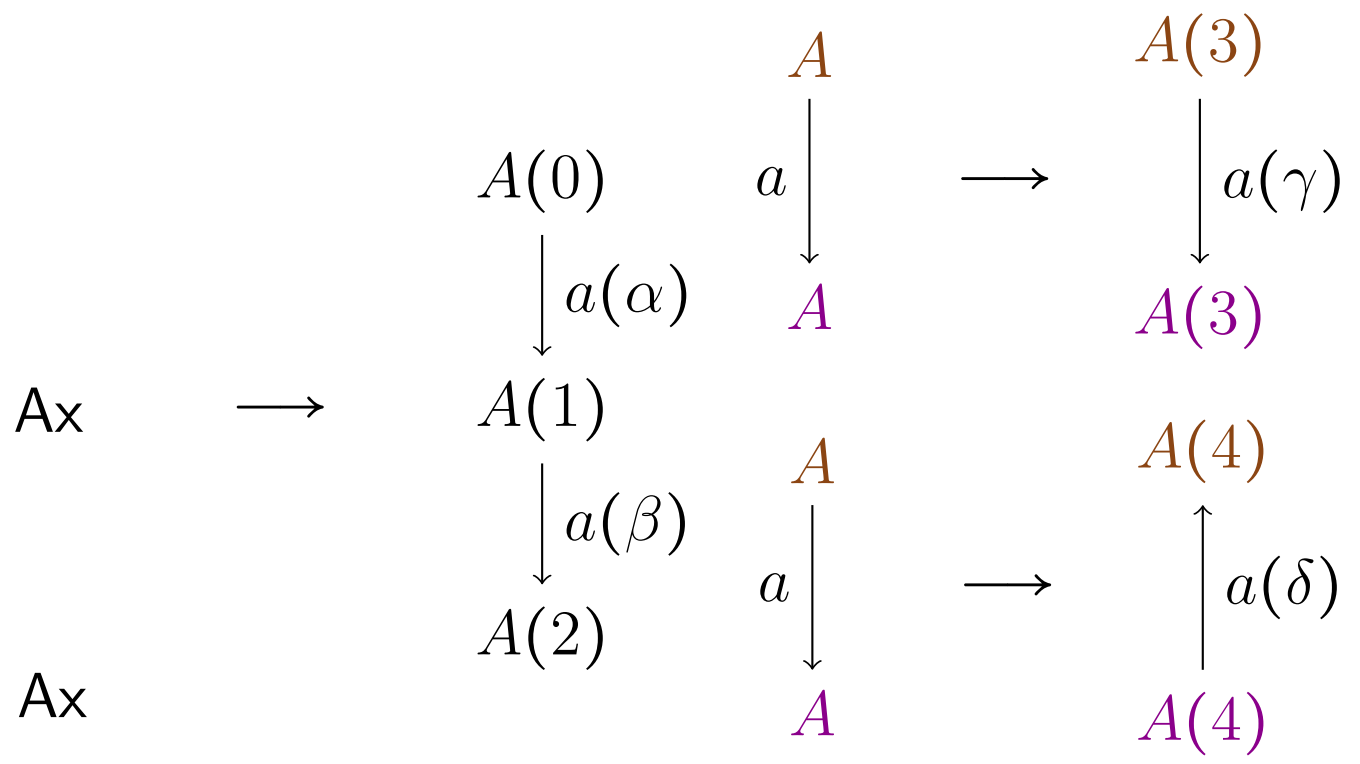

*The first two steps of possible derivations are depicted on the right-below.*

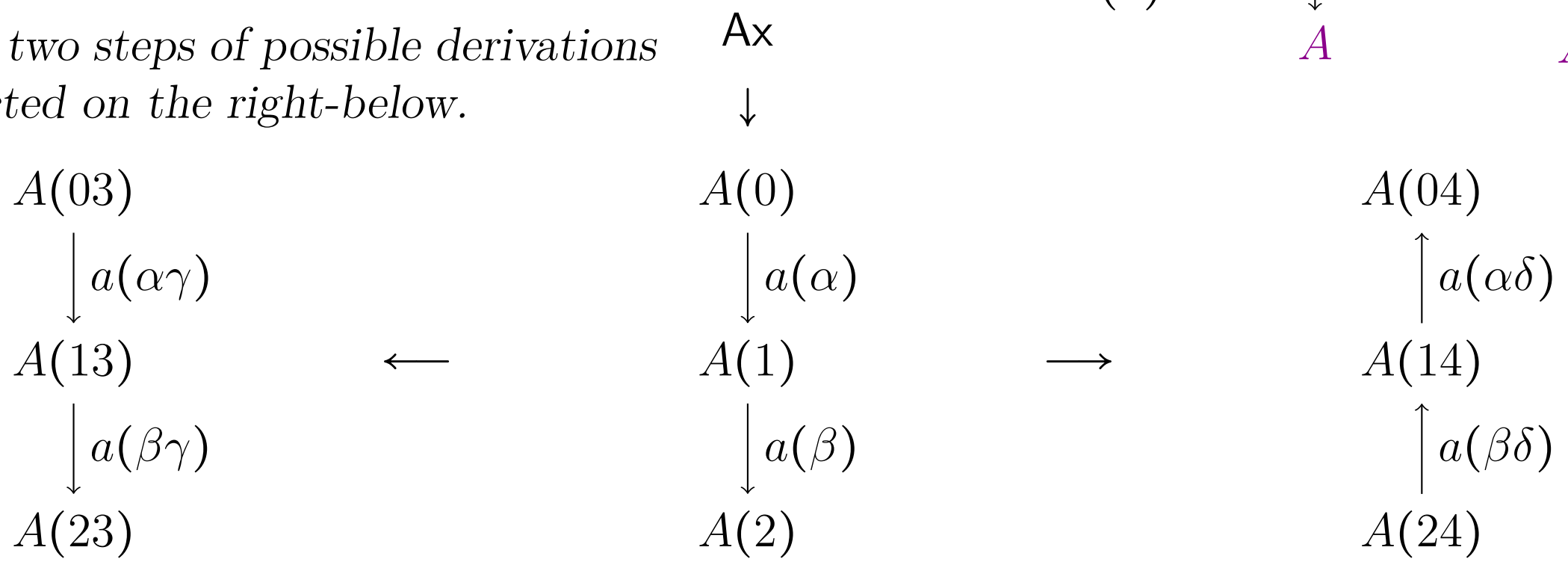

*Seen as a graph, this 0L graph grammar consists of three isolated vertices forming* V*, two of which are equipped with a loop, as depicted on the right. The additional information about* E *mentioned in the definition of 0L graph grammar is put in brackets.*

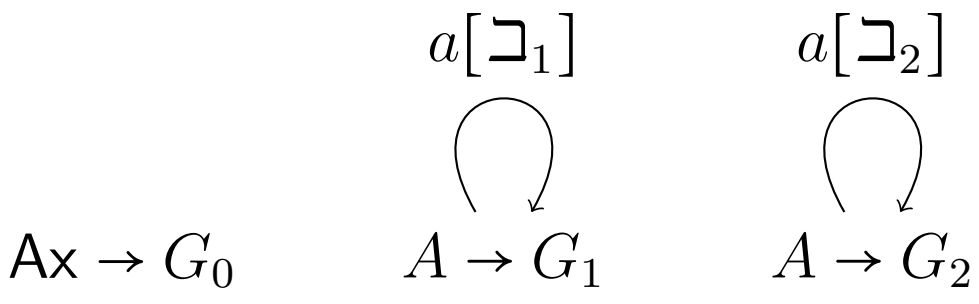

*This grammar is not complete. It lacks edges between the two above $A$-labelled vertices (in each direction). It is neither vD0L nor eD0L.*

**Example 4.4**

*By extending Example 4.3 with an additional edge production $(A \to G_1) \xrightarrow{a} (A \to G_2) \to \beth_3$ where $\beth_3 := \{\overrightarrow{a}(3,\zeta,4)\}$ which can be depicted as $A \xrightarrow{a} A \longrightarrow A(3) \xrightarrow{a(\zeta)} A(4)$, two more graphs derive from $G_0$ in a single step:* $A(03) \xrightarrow{a(\alpha\zeta)} A(14) \xleftarrow{a(\beta\delta)} A(24)$

$$A(03) \xrightarrow{a(\alpha\gamma)} A(13) \xrightarrow{a(\beta\zeta)} A(24) \ .$$

*Observe that these graph share their vertices with graphs derived in Example 4.3. Also, each new graph shares one edge with a graph from the latter example.*

**Example 4.5**

*D0L grammar generating a language of $2^n \times 2^n$ grids, for $n \geq 1$.*

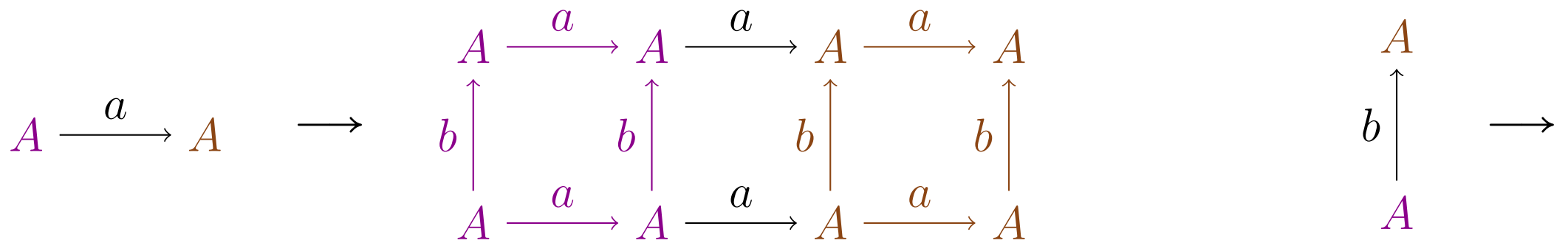

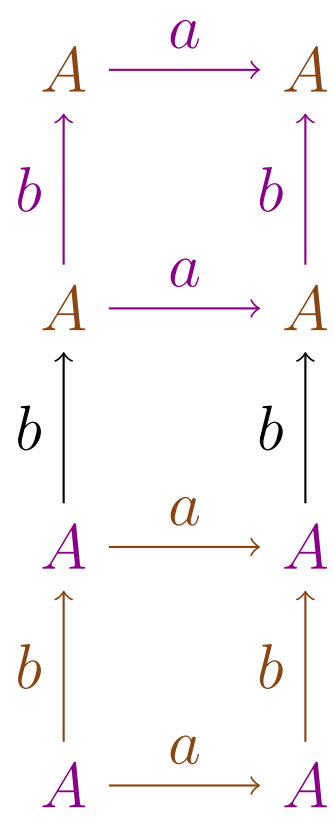

*Note that the unique vertex production is easily inferred from the two edge productions. As it is not explicitly specified, the axiom carries the unique vertex label $A$.*

**Example 4.6**

*D0L grammar for undirected unlabelled graphs (formally $|\Gamma| = |\Sigma| = 1$) generating all hypercubes via its unique derivation depicted below.*

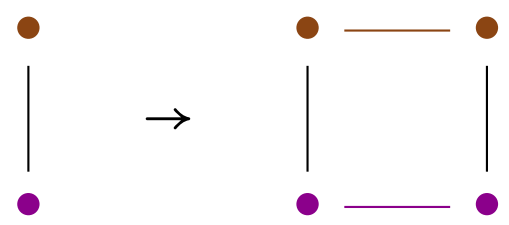

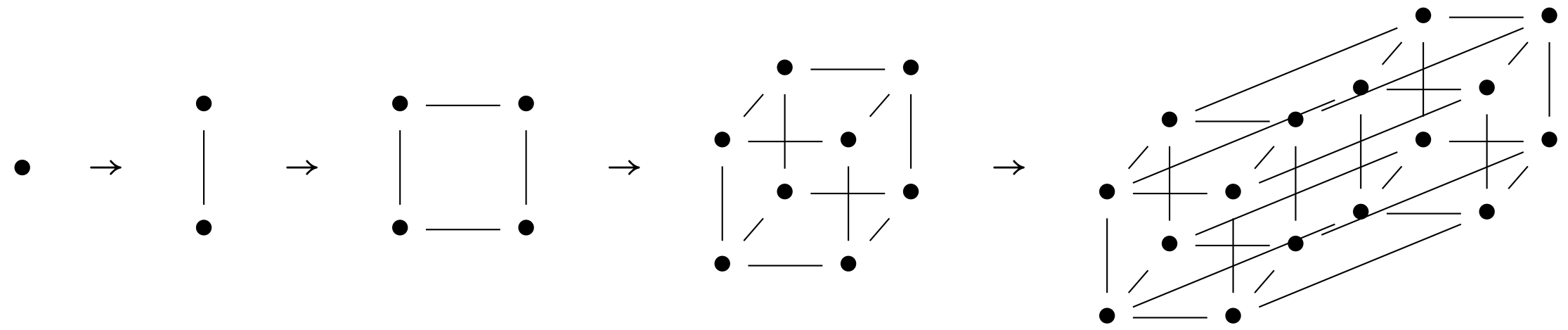

Observe that by definition, a 0L graph grammar cannot be ambiguous. Indeed, every graph in $\mathscr{L}(\mathcal{G})$ has a unique derivation. If, two distinct derivations lead to isomorphic graphs, say $(V_1, E_1, _, _)$ and $(V_2, E_2, _, _)$, then $V_1 \cup E_1 \neq V_2 \cup E_2$ (see Example 4.3). However, two distinct graphs need not to be disjoint as pointed out in Example 4.4.

As $\mathscr{L}(\mathcal{G})$ is in one-to-one correspondence with the set of derivations of $\mathcal{G}$, one can stratify $\mathscr{L}(\mathcal{G})$ according to derivation length. For $l \in \mathbb{N}$, an *$l$-layer of $\mathscr{L}(\mathcal{G})$* is the set, written $\mathscr{L}_l(\mathcal{G})$, of graphs having derivations of length $l$ according to $\mathcal{G}$. With that respect, the following two lemmas are as obvious as important.

**Lemma 4.7.** *$\mathscr{L}_l(\mathcal{G}) \subseteq \mathscr{G}_l(\Gamma, \Sigma, \Theta, \Omega)$, for every $l \in \mathbb{N}$ and every $\mathcal{G} \in 0\mathrm{L}(\Gamma, \Sigma, \Theta, \Omega)$.*

**Lemma 4.8.** *For every unix 0L graph grammar $\mathcal{G}$ and every $l \in \mathbb{N}$*
(1) *$|\mathscr{L}_l(\mathcal{G})| \leq 1$ and*
(2) *if $|\mathscr{L}(\mathcal{G})| = \omega$, then $|\mathscr{L}_l(\mathcal{G})| = 1$.*

We now turn to a few typical language-theoretic decision problems.

**Theorem 4.9.** *The following problems*

Instance: *0L graph grammar $\mathcal{G}$, label $\lambda$ of $\mathcal{G}$ and production $P$ of $\mathcal{G}$.*
Finiteness: *Is $\mathscr{L}(\mathcal{G})$ finite ?*
Emptiness: *$\mathscr{L}(\mathcal{G}) = \varnothing$ ?*
Accessibility: *Is $\lambda$ accessible in $\mathcal{G}$ ?*
Usefulness: *Is $P$ useful for $\mathcal{G}$ ?*

*are*

(i) *decidable for complete 0L graph grammars without non-terminals and for D0L graph grammars even with non-terminals,*
(ii) *undecidable for unix eD0L graph grammars, with non-terminals in the case of emptiness.*

For every 0L graph grammar $\mathcal{G}$, by enumerating $\mathscr{L}_l(\mathcal{G})$, one can decide if a concrete graph $G \in \mathscr{G}_l(\Gamma, \Sigma, \Theta, \Omega)$ is in $\mathscr{L}(\mathcal{G})$. The membership up to isomorphism considered in the next statement is a bit more challenging.

**Proposition 4.10.** *The (abstract) membership problem is decidable for D0L graph grammars even with non-terminals. It is undecidable for unix eD0L graph grammars.*

We let $\mathcal{V}(\mathcal{G}) := \bigcup_{(V,_,_,_) \in \mathscr{L}(\mathcal{G})} V$ (resp. $\mathcal{E}(\mathcal{G}) := \bigcup_{(_,E,_,_) \in \mathscr{L}(\mathcal{G})} E$, $\mathcal{I}(\mathcal{G}) := \bigcup_{(_,_,_,\gimel) \in \mathscr{L}(\mathcal{G})} \gimel$) denote the sets of all vertices (resp. edges, incidences) generated by $\mathcal{G}$. We write $\mathcal{V}(\mathcal{G})_N$, $\mathcal{E}(\mathcal{G})_N$ and $\mathcal{I}(\mathcal{G})_N$, when generating is restricted to $\mathscr{L}(\mathcal{G})_N$ for some set $N$ of non-terminals. The sets so defined are stratified, with layers $\mathcal{V}_l(\mathcal{G}) := \{v \in \mathcal{V}(\mathcal{G}) \ : \ |v| = l\}$, $\mathcal{E}_l(\mathcal{G}) := \{v \in \mathcal{E}(\mathcal{G}) \ : \ |v| = l\}$ and $\mathcal{I}_l(\mathcal{G}) := \{a(u, w, v) \in \mathcal{I}(\mathcal{G}) \ : \ |u| = |v| = |w| = l\}$ for $l \in \mathbb{N}$. Similarly for $\mathcal{V}_l(\mathcal{G})_N$, $\mathcal{E}_l(\mathcal{G})_N$ and $\mathcal{I}_l(\mathcal{G})_N$.

The *incompatibility relation (of $\mathcal{G}$)*, written "$\between$", is defined on $(\Theta \cup \Omega)^*$ as

$$\between \ := \ \between_0 \cup \between_1 \cup \between_1^{-1} \cup \between_2 \cup \between_2^{-1}$$

where

$$
\begin{aligned}
\mathrel{)(}_0 &:= \big\{(uis, ujt) \in (\mathcal{V}_l(\mathcal{G}) \cup \mathcal{E}_l(\mathcal{G}))^2 : l \in \mathbb{N}_+ \wedge \ \mathrm{lhs}(\mathrm{prd}(i)) = \mathrm{lhs}(\mathrm{prd}(j)) \wedge \mathrm{prd}(i) \neq \mathrm{prd}(j)\big\} \\
\mathrel{)(}_1 &:= \big\{(us, wt) \in (\mathcal{V}_l(\mathcal{G}) \cup \mathcal{E}_l(\mathcal{G}))^2 : l \in \mathbb{N}_+ \wedge \ \exists\, a \in \Sigma\ \exists\, u_1, u_2 \in \mathcal{V}(\mathcal{G})\ \big(a(u_1, w, u_2) \in \mathcal{I}(\mathcal{G}) \wedge \\
&\qquad (u \mathrel{)(}_0 u_1 \vee u \mathrel{)(}_0 u_2)\big)\big\} \\
\mathrel{)(}_2 &:= \big\{(ws, w't) \in \mathcal{E}_l(\mathcal{G})^2 : l \in \mathbb{N}_+ \wedge \ \exists\, a \in \Sigma\ \exists\, u_1, u_2 \in \mathcal{V}(\mathcal{G})\ \big(a(u_1, w, u_2) \in \mathcal{I}(\mathcal{G}) \wedge \\
&\qquad (u_1 \mathrel{)(}_1 w' \vee u_2 \mathrel{)(}_1 w')\big)\big\}
\end{aligned}
$$

The incompatibility relation is intended to capture the impossibility of two elements, vertices or edges, of the same layer, belonging to the same graph. Indeed, $\mathrel{)(}_0$ says that applying two different productions on the same vertex or edge results in two sets (of vertices or edges) that cannot belong to the same graph. A bit more subtle is the impossibility related to inherited edges as defined by $\mathrel{)(}_1$ and $\mathrel{)(}_2$. The former says that an inherited edge is incompatible with every vertex or edge which is incompatible with its source or its target. The latter says that two inherited edges are incompatible if one is incident to a vertex incompatible with the other. Note that $\mathrel{)(}$ is irreflexive and, as $\mathrel{)(}_0 = \mathrel{)(}_0^{-1}$, it is also symmetric. The layerwise complement of $\mathrel{)(}$ is the *compatibility relation*, written $\circ$ and defined by:

$$
\circ \ := \ \bigcup_{l \in \mathbb{N}} \big((\mathcal{V}_l(\mathcal{G}) \cup \mathcal{E}_l(\mathcal{G}))^2 \setminus \mathrel{)(}\big) \ .
$$

A set $S \subseteq \mathcal{V}_l(\mathcal{G}) \cup \mathcal{E}(\mathcal{G})$ is *compatible* (resp. *maximal compatible*) if its elements are pairwise compatible, viz., $S^2 \subseteq \circ$ (resp. and is $\subseteq$-maximal among compatible sets). Note that $\circ$ is reflexive and symmetric but not transitive. The relevance of this notion is highlighted by the following lemma closing this section.

**Lemma 4.11.** *If $\mathcal{G}$ is a complete 0L graph grammar, then*

$$
\{S \subseteq \mathcal{V}_l(\mathcal{G}) \cup \mathcal{E}_l(\mathcal{G}) : S \text{ is maximal compatible}\} \ = \ \{V \cup E : (V, E, _, _) \in \mathscr{L}_l(\mathcal{G})\} \, .
$$

# 5 Expanders

In this section we focus on families of efficiently connected graphs. Such families called expanders (see [19] and [26] for more details) have been studied mostly in the case of undirected graphs of uniform degree, also called $d$-regular, where all members of a family have its vertices of the same degree $d$. This section is written assuming such setting.

Intuitively, an expander is a large graph which is sparse but efficiently connected relatively to its degree. This paradigm can be formalised using the Cheeger constant also known as the isoperimetric number, written $h(G)$ for graph $G = (V, E)$ and defined by

$$
h(G) \ := \ \min\Big\{\frac{|\partial S|}{|S|} \ : \ S \subset V \ \wedge \ 0 < |S| \leq \tfrac{1}{2}|V|\Big\} \text{ where } \partial S \ := \ \big\{\{u, v\} \in E \ : \ u \in S \wedge v \in V \setminus S\big\} \ .
$$

An infinite family $\{G_i : i \in I\}$ is called *an expander family* if all its members have the same degree and there exists $\epsilon \in \mathbb{R}_+$ such that $h(G_i) > \epsilon$ for all $i \in I$. Although among graphs with $n$ vertices the maximum value of the Cheeger constant is obtained for clique $h(K_n) = n/2$, because the degree of cliques grows with their size, $\{K_n : n \in \mathbb{N}\}$ is not an expander family. As the problem of computing the Cheeger constant is NP-hard, one uses the isoperimetric inequalities of [1]

$$
\frac{d - \lambda_2}{2} \leq h(G) \leq \sqrt{2d(d - \lambda_2)}
$$

where $d$ is the degree of $G$ and $\lambda_2$ stands for the second largest eigenvalue of the normalised, by factor $1/d$, adjacency matrix of $G$. Because $\lambda_2 > 0$ one looks for graph families where all members have the second eigenvalue as small as possible. The smallest values are hit by *Ramanujan graphs* where $\lambda_2 \leq 2\sqrt{d-1}$. Efficient construction of expanders is a challenging research topic. In this section we review a few graph operations used in the most elementary combinatorial constructions of expanders and show how these can be implemented within 0L graph grammar framework. Algebraic constructions (extensively covered in survey [26]) where expanders occur as Cayley graphs are beyond the scope of this paper.

### 5.1 2-lifts

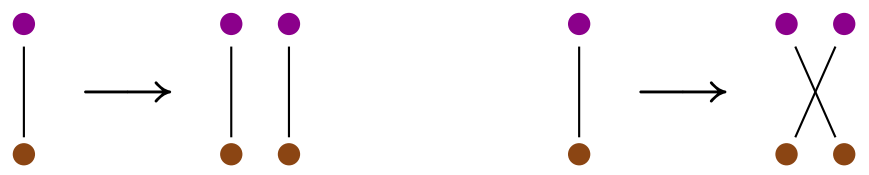

As established in [29] and [30], starting from an arbitrary bipartite Ramanujan graph, the above very simple two rules let generate an infinite family of bipartite Ramanujan graphs. These two edge productions together with vertex production • → • • and another one, which from the axiom yields the starting bipartite Ramanujan graph, say $B$, form a complete vD0L graph grammar, say $\mathcal{G}_B$. However, only a strict subset of $\mathscr{L}(\mathcal{G}_B)$ is Ramanujan. Indeed, it is shown in [29] that there exists a derivation along which encountered graphs form the desired family.[5] As it requires a nondeterministic choice among the above two vertex productions, the $n$-th graph of the family is obtained in time $o(2^n)$. However, an independently discovered essential improvement of this method leads to a polynomial time construction [6]. A family obtained is not limited to $d$-regular graphs. Indeed, the construction lets derive a family of $(c,d)$-biregular Ramanujan graphs starting from one such graph. A bipartite graph is $(c,d)$-biregular if all vertices in one part have degree $c$ and those in the remaining part have degree $d$.

### 5.2 Shift 4-lifts

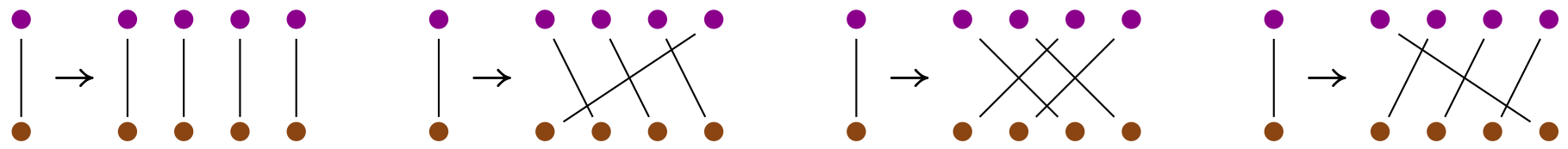

The construction using 2-lifts is further improved in [5] by using shift 4-lifts (or shift 3-lifts defined similarly). This operation can be written as the above four edge productions and one vertex production • → • • • • . Together with an axiom, we get a complete vD0L graph grammar. As in the example of 2-lifts, by selecting an appropriate derivation, one obtains an infinite family of bipartite Ramanujan graphs, provided that from the axiom one starts with its smallest member.

### 5.3 Graph operations using rotation maps

In this subsection we review graph operations introduced in [34] together with rotation maps. The latter equip an undirected graph with an additional structure: every vertex has its numbering of incident edges. Thus, every edge is equipped with two numbers. Such an enriched edge between vertices $u$ and $v$ can be written $\{\langle u,i\rangle,\langle v,j\rangle\}$, where $i,j \in [d]$ for a $d$-regular graph, to mean that $v$ is the $i$-th neighbour of $u$ and $u$ is the $j$-th neighbour of $v$. In [34] it is denoted by $\mathrm{Rot}(u,i) = (v,j)$, or, equivalently, by $\mathrm{Rot}(v,j) = (u,i)$. To make it sound, one requires that for all enriched edges $\{\langle u,i\rangle,\langle v,l\rangle\}$ and $\{\langle u,j\rangle,\langle w,k\rangle\}$, if $i = j$ then $v = w$ and $l = k$. Note that with this notation multiple edges between a pair of vertices arise naturally without a need of declaring an explicit set of edges. Rotation maps can be drawn using undirected edges with labels at each extremity, $u \overset{i\ \ j}{\text{---}} v$, and implemented in a 0L graph grammar by two opposite directed incidences with a pair of labels drawn as $u \xrightarrow{(i,j)} v$ and $v \xrightarrow{(j,i)} u$. One might suggest simplifying the numbering of the neighbours of each vertex so that $i = j$ for every enriched edge $\{\langle u,i\rangle,\langle v,j\rangle\}$ and represent it as $u \overset{i}{\text{---}} v$ but this is not possible in general as it can be quickly checked on $K_3$.

For the reminder of this section $\mathscr{G}(V,d,\lambda)$ stands for the set of $d$-regular graphs with vertices $V$ and the second largest eigenvalue of the normalised adjacency matrix at most $\lambda$. The set of enriched edges of graph $G$ is written $\mathcal{E}(G)$.

[5]The original statement of [29] does not use graph grammars terminology.

### Replacement product

The replacement product takes two graphs $G_1 \in \mathscr{G}([n_1], d_1, \lambda_1)$ and $G_2 \in \mathscr{G}([d_1], d_2, \lambda_2)$ and outputs $G_1 ⓡ\, G_2 \in \mathscr{G}\big([n_1] \times [d_1], d_2 + 1, \lambda\big)$ with the following bound on $\lambda$ as established in [34]:

$$\lambda \quad \leq \quad \Big(\frac{d_2^3}{d_2^3 + 1} + \big(1 - \frac{d_2^3}{d_2^3 + 1}\big)\Big(\frac{1}{2}(1 - \lambda_2^2)\lambda_1 + \frac{1}{2}\sqrt{(1 - \lambda_2^2)\lambda_1^2 + 4\lambda_2^2}\,\Big)\Big)^{\frac{1}{3}} \ .$$

Enriched edges defining $G_1 ⓡ\, G_2$ are

$$\begin{aligned} \mathcal{E}(G_1 ⓡ\, G_2) \quad &:= \quad \big\{\{((u,k),i\rangle, ((u,l),j\rangle\} \ : \ u \in [n_1] \wedge \{(k,i\rangle, (l,j\rangle\} \in \mathcal{E}(G_2)\big\} \\ &\qquad \cup \ \big\{\{((u,k), d_2 + 1\rangle, ((v,l), d_2 + 1\rangle\} \ : \ \{(u,k\rangle, (v,l\rangle\} \in \mathcal{E}(G_1)\big\} \ . \end{aligned}$$

For all graphs $G_0 \in \mathscr{G}(_, d + 1, _)$ and $H \in \mathscr{G}([d + 1], d, _)$ we can write D0L graph grammar $\mathcal{G}_{ⓡH,G_0}$ with axiom production $\mathsf{A} \to G_0$, another vertex production $\bullet \to H$ and the set of edge productions $\{P_{k,l} : k, l \in [d + 1]\}$ where each $P_{k,l}$ has the following form

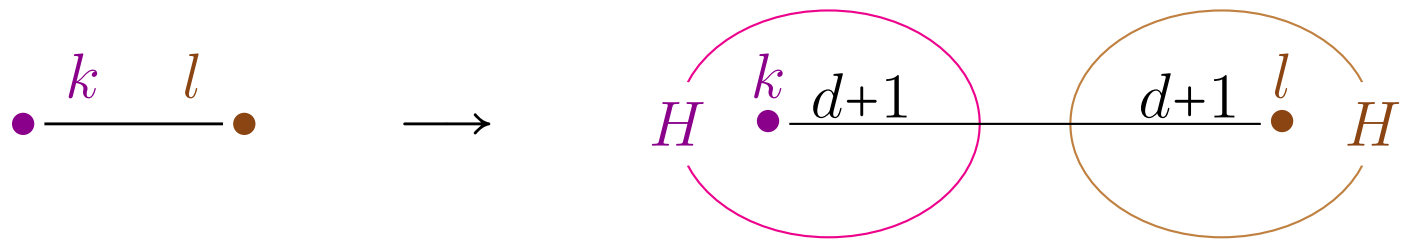

Note that $k$ and $l$ label the edge of lhs$(P_{k,l})$ whereas in either copy of $H$ in $P_{k,l}$ they name its vertices. This grammar rewrites any graph $G \in \mathscr{G}(_, d + 1, _)$ into $G$ ⓡ $H \in \mathscr{G}(_, d + 1, _)$. When $G_0$ is a good expander and $H$ is an odd cycle, $\mathscr{L}(\mathcal{G}_{ⓡH,G_0})$ is an expander family of degree three [34]. However, in general, the expansion due to replacement product is not always satisfactory.

### Balanced replacement product

A slight improvement of expansion is obtained with the balanced replacement product. It takes two graphs $G_1 \in \mathscr{G}([n_1], d_1, \lambda_1)$ and $G_2 \in \mathscr{G}([d_1], d_2, \lambda_2)$ and outputs $G_1 ⓑ\, G_2 \in \mathscr{G}\big([n_1] \times [d_1], 2d_2, \lambda\big)$ with the following bound on $\lambda$ as established in [34]:

$$\lambda \quad \leq \quad \Big(\frac{7}{8} + \frac{1}{16}(1 - \lambda_2^2)\lambda_1 + \frac{1}{16}\sqrt{(1 - \lambda_2^2)\lambda_1^2 + 4\lambda_2^2}\,\Big)^{\frac{1}{3}} \ .$$

Enriched edges defining $G_1 ⓑ\, G_2$ are

$$\begin{aligned} \mathcal{E}(G_1 ⓑ\, G_2) \quad &:= \quad \big\{\{((u,k),i\rangle, ((u,l),j\rangle\} \ : \ u \in [n_1] \wedge \{(k,i\rangle, (l,j\rangle\} \in \mathcal{E}(G_2)\big\} \\ &\qquad \cup \ \big\{\{((u,k),i\rangle, ((v,l),i\rangle\} \ : \ \{(u,k\rangle, (v,l\rangle\} \in \mathcal{E}(G_1) \wedge d_2 < i \leq 2d_2\big\} \ . \end{aligned}$$

Observe that the balanced replacement product differs from the former replacement product only by the multiplicity of edges between two copies of $G_2$ which is 1 in the former and $d_2$ in the present product. Thus, a vertex has equal number of neighbours within and outside each copy of $G_2$. The construction of D0L graph grammars is similar to those from the latter subsection except for multiple edges. For all graphs $G_0 \in \mathscr{G}(_, 2d, _)$ and $H \in \mathscr{G}([2d], d, _)$, one can write D0L graph grammar $\mathcal{G}_{ⓑH,G_0}$. Expander families are obtained as for the former replacement product. In particular, the degree is kept constant because any graph $G \in \mathscr{G}(_, 2d, _)$ rewrites in a single step into $G$ ⓑ $H \in \mathscr{G}(_, 2d, _)$.

### Zig-zag product

An even better expansion is obtained with the zig-zag product. It takes two graphs $G_1 \in \mathscr{G}([n_1], d_1, \lambda_1)$ and $G_2 \in \mathscr{G}([d_1], d_2, \lambda_2)$ and outputs $G_1 ⓩ\, G_2 \in \mathscr{G}\big([n_1] \times [d_1], d_2^2, \lambda\big)$ with the following bound on $\lambda$ as established in [34]:

$$\lambda \quad \leq \quad \Big(\frac{1}{2}(1 - \lambda_2^2)\lambda_1 + \frac{1}{2}\sqrt{(1 - \lambda_2^2)\lambda_1^2 + 4\lambda_2^2}\,\Big)^{\frac{1}{3}} \ .$$

The "number" of a neighbour in $G_1 ⓩ G_2$ is a pair in $[d_2]^2$ according to the original definition. Enriched edges defining $G_1 ⓩ G_2$ are

$$\mathcal{E}(G_1 ⓩ G_2) := \Big\{\{((u,k),(i,j)\rangle,((v,l),(j',i')\rangle\} \; : \; \exists\, k',l' \in [d_1] \quad \big(\{(k,i\rangle,(k',i'\rangle\},\{(l,j'\rangle,(l',j\rangle\} \in \mathcal{E}(G_2) \wedge \{(u,k'\rangle,(v,l'\rangle\} \in \mathcal{E}(G_1)\big)\Big\} \; .$$

An enriched edge $u \frac{k' \quad l'}{\quad} v$ of $G_1$ gives raise to $d_2^2$ such new enriched edges between copies $\{u\}\times V_2$ and $\{v\}\times V_2$ of vertices $V_2$ of $G_2$. Label $k'$ (resp. $l'$) determines the vertex that is "the entry point" of copy $\{u\}\times V_2$ (resp. $\{v\}\times V_2$). There is an edge between vertex $(u,k)$ in copy $\{u\}\times V_2$ and vertex $(v,l)$ in copy $\{v\}\times V_2$ if one can take an edge $k \frac{i \quad i'}{\quad} k'$ in $G_2$ from $k$ to entry point $k'$ and an edge $l \frac{j' \quad j}{\quad} l'$ in $G_2$ from $l$ to entry point $l'$. Because of edge $u \frac{k' \quad l'}{\quad} v$ in $G_1$, these entry points allow to "jump" between copies $\{u\}\times V_2$ and $\{v\}\times V_2$. The label on $(u,k)$'s side of created enriched edge $(u,k) \frac{(i,j) \quad (j',i')}{\quad} (v,l)$ describes the corresponding "route with jump" from $(u,k)$ to $(v,l)$ as follows. *From $k$ take direction $i$ to get to the entry point of current copy. Jump and you land at the entry point of $v$'s copy. From there, take direction $j$ and you are there.* Label $(j',i')$ on $(v,l)$'s side describes the same route in the reverse direction.

In order to implement the zig-zag product in D0L graph grammar so that it can be iterated, the set of labels used for neighbours numbering must be kept constant at every derivation step. The choice we made is to take, as the right hand side of the axiom production, $G_0 \in \mathscr{G}(_, d^2, _)$, where each edge has at either extremity a pair in $[d]^2$ as label. With $H \in \mathscr{G}([d]^2, d, _)$, one can complete the construction of D0L graph grammar $\mathcal{G}_{ⓩH,G_0}$ suitable for such iteration. Besides the axiom, $\mathcal{G}_{ⓩH,G_0}$ has one vertex production $\bullet \to V$ where $V$ is the edgeless version of $H$. The set of edge productions $\{P_{\overline{k}',\overline{l}'} : \overline{k}',\overline{l}' \in [d]^2\}$ is defined so that each lhs$(P_{\overline{k}',\overline{l}'})$ is $\bullet \frac{\overline{k}' \quad \overline{l}'}{\quad} \bullet$ and the corresponding rhs$(P_{\overline{k}',\overline{l}'})$ creates edges between each pair $\{\overline{k},\overline{l}\}$ of neighbours of $\overline{k}'$ and $\overline{l}'$ in their respective copies of $V$. Every such edge of rhs$(P_{\overline{k}',\overline{l}'})$ can be drawn as without dashed edges which serve here only as explanation. Indeed, the latter represent original edges of $H$ which however do not appear in $\mathcal{G}_{ⓩH,G_0}$ as it only uses edgeless version of $H$. Remember that $\overline{k},\overline{l},\overline{k}',\overline{l}'$ are pairs in $[d]^2$ whereas $i,j,i',j'$ are numbers in $[d]$.

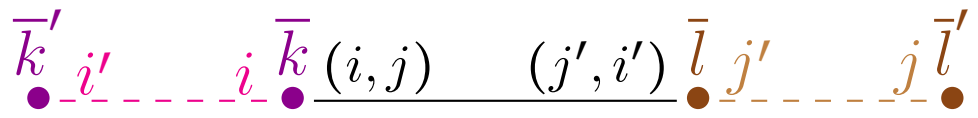

**Example 5.1**

*In this example $d = 2$. We use $\{0,1\}$ for neighbours numbering in $H$ (depicted on the right) and $\{0,1\}^2$ for naming its vertices and for neighbours numbering of $G_0 := K_5$. We have the following edge productions that apply on $K_5$ below.*

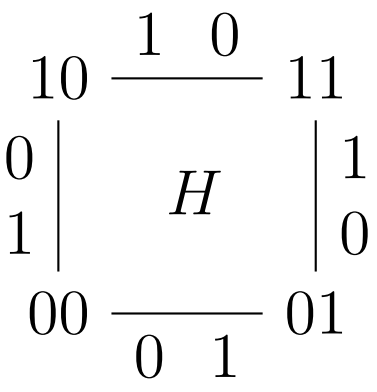

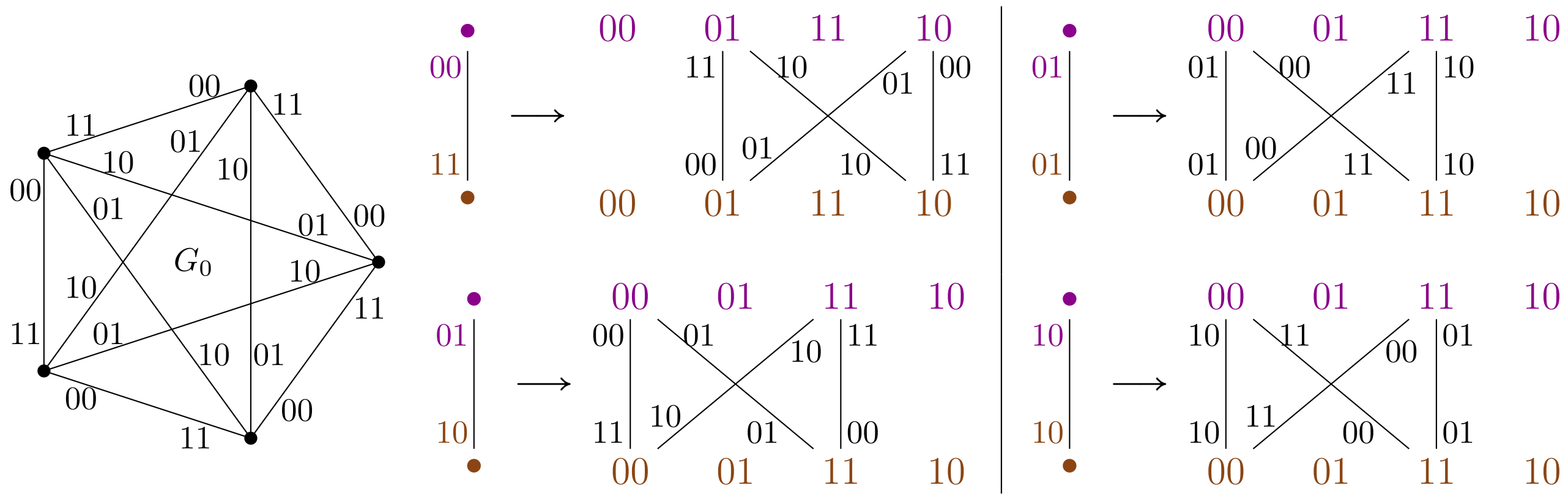

*For more clarity, the names of vertices are written in the right hand sides of edge productions. Although the number of possible edge labellings is $\big|\{\{ab,cd\} : a,b,c,d \in \{0,1\}\}\big| = 10$, but only*

*four appear in $G_0$ and these form the left hand sides of the above edge productions. Moreover, their right hand sides only introduce three types of enriched edges among those of $G_0$. Thus, this D0L graph grammar is like $\mathcal{G}_{ⓩ H,G_0}$ except that it has no useless productions. Consequently the language generated here is as expected $\{G_n : n \in \mathbb{N}\}$ where $G_{i+1} = G_i ⓩ H$.*

To close this section, we need to mention that the most efficient construction of expanders of [34] with the zig-zag product requires an additional operation of squaring. It consists in replacing an incident pair of edges $u$—$v$—$w$ by a single edge $u$—$w$. Within Lindenmayer graph grammars, squaring can be handled in 1L (instead of 0L) framework, unfortunately, at the expense of most decidability results.

# 6 0L graph languages as models

We start this section by fixing first order language $\mathrm{FO}(\Gamma \cup \Sigma)$ and its extension $\mathrm{FO}(\Upsilon)$ for expressing properties or queries concerning 0L graph languages. Before focusing on $\mathrm{FO}(\Gamma \cup \Sigma)$, questions one would like to ask are conveniently qualified in a more general scope than graph languages, namely that of arbitrary family of structures $\mathscr{S}$ together with adequate logic $\mathcal{L}$. We consider the

LANGUAGE $\mathcal{L}$-CHECKING PROBLEM FOR FAMILY $\mathscr{S}$ OF STRUCTURES
Instance: a countable set of countable structures $\mathcal{S} \in \mathscr{S}$ and an $\mathcal{L}$-sentence $\varphi$
Question: $\exists S \in \mathcal{S}\ S \models \varphi$?

Replacing the existential quantifier in the question by the universal one leads to an equivalent checking problem. Consequently, any finite Boolean combination of existential and universal questions leads to an equivalent checking problem too. The reader should note the difference of the above problems with the well-known

MODEL $\mathcal{L}$-CHECKING PROBLEM FOR CLASS $\mathcal{S}$ OF STRUCTURES
Instance: a countable structure $S \in \mathcal{S}$ and an $\mathcal{L}$-sentence $\varphi$ (resp. $\mathcal{L}$-formula $\varphi(\mathcal{X})$),
Question: $S \models \varphi$?

Whereas the instances in the latter problem range over individual structures, they range over sets of structures in the former. Both can be confused when the set of sentences is, up to logical equivalence, in one-to-one correspondence with family $\mathscr{S}$, like for instance in the case of MSO and the family of regular sets.

In the present paper, we are interested in the following case of the former generic problems: *the language* $\mathrm{FO}(\Gamma \cup \Sigma)$*-checking* (resp. $\mathrm{FO}(\Gamma \cup \Sigma)$*-querying*)*problem for the family of* 0*L graph languages*. Given a 0L grammar $\mathcal{G}$ and an $\mathrm{FO}(\Gamma \cup \Sigma)$ sentence $\varphi$, we ask whether a graph in $\mathscr{L}(\mathcal{G})$ satisfies $\varphi$. Sentence $\varphi$ can for instance express the property that a graph has a clique of a given fixed size as induced subgraph. To establish the decidability of the above problem, we proceed by reduction of the language $\mathrm{FO}(\Gamma \cup \Sigma)$-checking problem for the family of 0L graph languages to the model $\mathrm{FO}(\Upsilon)$-checking problem for a class of infinite structures associated with those languages, where the sentences are written using relational signature $\Upsilon$ extending $\Gamma \cup \Sigma$. By representing $\mathscr{L}(\mathcal{G})$ as an adequate single structure, say $\mathcal{G}^{\uparrow}$, given 0L graph grammar $\mathcal{G}$ and sentence $\varphi \in \mathrm{FO}(\Gamma \cup \Sigma)$, under certain assumptions, one can reduce a question "$\exists G \in \mathscr{L}(\mathcal{G})\ G \models \varphi$?" to a question "$\mathcal{G}^{\uparrow} \models \psi$?" with $\psi \in \mathrm{FO}(\Upsilon)$. In doing so, one needs to separate somehow individual graphs within $\mathcal{G}^{\uparrow}$ using an appropriate sentence $\psi$ associated to $\varphi$. Indeed, as pointed out in Example 4.4. two graphs in $\mathscr{L}(\mathcal{G})$ need not be disjoint.

Variables from a countable set $\mathscr{X}$ are used for writing $\mathrm{FO}(\Gamma \cup \Sigma)$ formulae expressing properties of a graph in $\mathscr{G}(\Gamma, \Sigma, _, _)$. Atomic formulae of $\mathrm{FO}(\Gamma \cup \Sigma)$ are

- "$x$ is an $A$-labelled vertex": $A(x)$, for $A \in \Gamma$ and $x \in \mathscr{X}$,

- "$y$ is an $a$-labelled edge from vertex $x_1$ to vertex $x_2$": $a(x_1, y, x_2)$, for $a \in \Sigma$, $x_1, y, x_2 \in \mathscr{X}$,
- equalities $x = y$, for $x, y \in \mathscr{X}$,
- and those for discriminating vertices and edges: $\mathsf{vrt}(x)$, $\mathsf{edg}(x)$, for $x \in \mathscr{X}$.

The closure of the atomic formulae under connectives $\neg$, $\supset$, $\vee$, and quantifiers $\exists$, $\forall$ yields the set of $\mathrm{FO}(\Gamma \cup \Sigma)$ formulae.

Additional symbols in $\mathrm{FO}(\Upsilon) \setminus (\Gamma \cup \Sigma)$ are used for writing the following atomic formulae:

- "vertex (resp. edge) $y$ is in a graph which derives from vertex (resp. or edge) $x$": $x \sqsubseteq y$ for $x, y \in \mathscr{X}$,
- "vertices or edges $x$ and $y$ are compatible": $x \circ y$ for $x, y \in \mathscr{X}$,
- "vertices or edges $x$ and $y$ belong to graphs of the same layer" $x \triangleq y$ for $x, y \in \mathscr{X}$.

The structure, written $\mathcal{G}^{\uparrow}$, underlying the language $\mathscr{L}(\mathcal{G})$ of 0L graph grammar $\mathcal{G}$ is a relational structure over signature $\Upsilon$. $\mathcal{G}^{\uparrow}$ consists of the universe of vertices and edges $\mathcal{V}(\mathcal{G}) \cup \mathcal{E}(\mathcal{G})$ where predicate symbols of $\Upsilon$ are interpreted in the following way:

- $A \in \Gamma$ as unary relation
$$\mathcal{G}^{\uparrow}(A) = \{v \in \mathcal{V}(\mathcal{G}) \ : \ \exists (V, _, \ell, _) \in \mathscr{L}(\mathcal{G}) \ \ell(v) = A\},$$
- $a \in \Sigma$, as ternary relation
$$\mathcal{G}^{\uparrow}(a) = \{(u, w, v) \ : \ \exists (_, _, _, \daleth) \in \mathscr{L}(\mathcal{G}) \ \ a(u, w, v) \in \daleth\},$$
- "$\circ$", as the compatibility relation defined in Sect. 4,
- "$\sqsubseteq$", as prefix ordering on $\mathcal{V}(\mathcal{G}) \cup \mathcal{E}(\mathcal{G})$,
- $\mathcal{G}^{\uparrow}(\triangleq) = \bigcup_{l \in \mathbb{N}} (\mathcal{V}_l(\mathcal{G}) \cup \mathcal{E}_l(\mathcal{G}))^2$, and,
- $\mathcal{G}^{\uparrow}(\mathsf{vrt}) = \mathcal{V}(\mathcal{G})$ and $\mathcal{G}^{\uparrow}(\mathsf{edg}) = \mathcal{E}(\mathcal{G})$.

For a set $N$ of non-terminals, $\mathcal{G}^{\uparrow \setminus N}$ is a restriction of $\mathcal{G}^{\uparrow}$ onto $\mathcal{V}(\mathcal{G})_N \cup \mathcal{E}(\mathcal{G})_N$.

In Sect. 7 we give a construction that yields an automatic presentation of $\mathcal{G}^{\uparrow}$ for every complete 0L graph grammar $\mathcal{G}$. Our construction brings the following result.

**Theorem 6.1.** *$\mathcal{G}^{\uparrow}$ (resp. $\mathcal{G}^{\uparrow \setminus N}$) has a decidable $\mathrm{FO}(\Upsilon)$ theory for every 0L graph grammar $\mathcal{G}$ which is complete (resp. deterministic with non-terminals $N$).*

The following theorem is essential for $\mathrm{FO}(\Gamma \cup \Sigma)$-language checking.

**Theorem 6.2.** *There is a translation $\tau\colon \mathrm{FO}(\Gamma \cup \Sigma) \to \mathrm{FO}(\Gamma \cup \Sigma \cup \{\triangleq\})$ in time $\mathcal{O}(nk + m)$, where $m$ is the length of the input formula, $n$ is its alternation rank and $k$ is the number of its variables, such that for every unix 0L graph grammar $\mathcal{G} \in 0\mathrm{L}(\Gamma, \Sigma, _, _)$ and every formula $\varphi(\mathfrak{X}) \in \mathrm{FO}(\Gamma \cup \Sigma)$, one has*
$$\mathcal{G}^{\uparrow}(\tau(\varphi(\mathfrak{X})) = \bigcup_{G \in \mathscr{L}(\mathcal{G})} G(\varphi(\mathfrak{X})) \ .$$
*In particular, when $\mathfrak{X} = \varnothing$, there exists $G \in \mathscr{L}(\mathcal{G})$ such that $G \models \varphi$, if, and only if, $\mathcal{G}^{\uparrow} \models \tau(\varphi)$.*

Additional translation steps can be done using the following proposition.

**Proposition 6.3.** *For every set of non-terminals $N \subseteq \Gamma \cup \Sigma$, there is a translation $\tau$ of $\mathrm{FO}(\Gamma \cup \Sigma)$ in time $\mathcal{O}(|N|)$, such that for every graph grammar $\mathcal{G} \in 0\mathrm{L}(\Gamma, \Sigma, _, _)$ and every formula $\varphi(\mathfrak{X}) \in \mathrm{FO}(\Gamma \cup \Sigma)$, one has $G(\tau(\varphi(\mathfrak{X}))) = G(\varphi(\mathfrak{X}))$ for every $G \in \mathscr{L}(\mathcal{G})_N$ and $G(\tau(\varphi(\mathfrak{X}))) = \varnothing$ for every $G \in \mathscr{L}(\mathcal{G}) \setminus \mathscr{L}(\mathcal{G})_N$.*
*In particular, when $\mathfrak{X} = \varnothing$, there exists $G_N \in \mathscr{L}(\mathcal{G})_N$ such that $G_N \models \varphi$, if, and only if, there exists $G \in \mathscr{L}(\mathcal{G})$ such that $G \models \tau(\varphi)$.*

*Proof.* The translation if obvious: $\tau(\varphi(\mathfrak{X})) = \varphi(\mathfrak{X}) \wedge \neg \exists x \bigvee_{A \in N \cap \Gamma} A(x) \wedge \neg \exists x y z \bigvee_{a \in N \cap \Sigma} a(x, y, z)$. □

Applied to Prop. 4.2, the above proposition allows for extending the translation of Thm. 6.2. Then, using Thm. 6.1, we conclude this section with the following result.

**Corollary 6.4.** $\mathrm{FO}(\Gamma \cup \Sigma)$*-language checking problem is decidable D0L graph grammars even with non-terminals.*

It is important to note that the corollary does not extend to unix graph grammars. The essential step of the proof uses Prop. 4.2 for transforming the input grammar into a complete grammar so as to benefit from Thm. 6.1. However, the grammar resulting from that step is not unix in general.

# 7 Automatic presentation of Lindenmayer languages

In this section we build an injective automatic presentation of $\mathcal{G}^{\uparrow}$, where the corresponding injection is in fact an identity map. The reader is supposed to be familiar with standard constructions on finite automata such as Boolean operations, concatenation or projection. The following additional constructions are used in the sequel.

- For a tuple automaton $\mathscr{B}$ we denote by $\mathscr{B}^{=}$ the tuple automaton which differs from $\mathscr{B}$ by labelling of transitions: each transition $p \xrightarrow{\overline{a}} q$ of $\mathscr{B}$ is replaced in $\mathscr{B}^{=}$ by $p \xrightarrow{\overline{a}\overline{a}} q$. Thus $[\![\mathscr{B}^{=}]\!] = \{(\overline{u}, \overline{u}) : \overline{u} \in [\![\mathscr{B}]\!]\}$ with $\mathsf{ar}(\mathscr{B}^{=}) = 2\mathsf{ar}(\mathscr{B})$.
- The *loose product* of tuple automata $\mathscr{A}_1 = (\mathbb{Q}_1, \Delta_1, \mathbb{I}_1, \mathbb{F}_1)$ and $\mathscr{A}_2 = (\mathbb{Q}_2, \Delta_2, \mathbb{I}_2, \mathbb{F}_2)$ is $\mathscr{A}_1 \times \mathscr{A}_2 = (\mathbb{Q}_1 \times \mathbb{Q}_2, \Delta_1 \times \Delta_2, \mathbb{I}_1 \times \mathbb{I}_2, \mathbb{F}_1 \times \mathbb{F}_2)$ where
$$\Delta_1 \times \Delta_2 \coloneqq \{(p,q) \xrightarrow{\overline{a}\overline{b}} (p',q') : p \xrightarrow{\overline{a}} p' \in \Delta_1 \wedge q \xrightarrow{\overline{b}} q' \in \Delta_2\}.$$
Thus $\mathscr{L}(\mathscr{A}_1 \times \mathscr{A}_2) = \bigcup_{l \in \mathbb{N}} \mathscr{L}_l(\mathscr{A}_1) \times \mathscr{L}_l(\mathscr{A}_2)$.
- For $\mathscr{A}_1$ and $\mathscr{A}_2$ as above and $\Delta \subseteq \mathbb{Q}_1 \times \Pi \times \mathbb{Q}_2$, where $\Pi$ is an alphabet, the *concatenation of* $\mathscr{A}_1$ *and* $\mathscr{A}_2$ *via* $\Delta$, written $\mathscr{A}_1 \xrightarrow{\Delta} \mathscr{A}_2$ is the automaton $(\mathbb{Q}_1 \cup \mathbb{Q}_2, \Delta_1 \cup \Delta \cup \Delta_2, \mathbb{I}_1, \mathbb{F}_2)$.

Each predicate symbol $\varrho$ of the signature of $\mathcal{G}^{\uparrow}$ has the corresponding automaton $\mathscr{A}(\varrho)$ of arity $\mathsf{ar}(\mathscr{A}(\varrho)) = \mathsf{ar}(\varrho)$ given as quadruple $\mathscr{A}(\varrho) = (\mathbb{Q}_\varrho, \Delta_\varrho, \iota_\varrho, \mathbb{F}_\varrho)$, where $\mathbb{Q}_\varrho$ is a finite set of states, $\Delta_\varrho \subseteq \mathbb{Q}_\varrho \times (\Theta \cup \Omega)^{\mathsf{ar}(\varrho)} \times \mathbb{Q}_\varrho$ is a transition relation, $\iota_\varrho$ is the initial state of $\mathscr{A}(\varrho)$ and $\mathbb{F}_\varrho \subseteq \mathbb{Q}_\varrho$ is its set of final states. The relevant predicate symbols are

- the equality,
- unary $\mathsf{vrt}$ and $\mathsf{edg}$,
- binary $\circlearrowleft$, $\hat{=}$ and $\sqsubseteq$.
- $A$, with $\mathsf{ar}(A) = 1$, for $A \in \Gamma$,
- $a$, with $\mathsf{ar}(a) = 3$, for $a \in \Sigma$,

Automata for $\hat{=}$, $\sqsubseteq$ and for the equality are obvious. Besides direct constructions, we also consider automata for FO-definable predicates without providing their construction. Indeed, such automata can be constructed in standard way from the corresponding formula as query automata (see Subsect. 3).

Here are the details of main automata, starting with vertex related ones.

- $\mathscr{A}(\mathsf{vrt}) \coloneqq (\mathbb{Q}_{\mathsf{vrt}}, \Delta_{\mathsf{vrt}}, \iota_{\mathsf{vrt}}, \mathbb{F}_{\mathsf{vrt}})$ is a deterministic automaton on $\Theta$ where

  $\mathbb{Q}_{\mathsf{vrt}} \coloneqq \Theta \cup \{\iota_{\mathsf{vrt}}\}$ is its set states,

  $\iota_{\mathsf{vrt}} \notin \Theta$ is its initial state,

  $\mathbb{F}_{\mathsf{vrt}} \coloneqq \Theta$ is its set of final states,

  $\Delta_{\mathsf{vrt}} \subseteq \mathbb{Q}_{\mathsf{vrt}} \times \Theta \times \mathbb{Q}_{\mathsf{vrt}}$ is its set of transitions defined by
  $$\Delta_{\mathsf{vrt}} \coloneqq \{i \xrightarrow{j} j \;:\; \mathrm{prd}(j) = \ell(i) \to _\} \cup \{\iota_{\mathsf{vrt}} \xrightarrow{j} j \;:\; \mathrm{prd}(j) = \mathsf{A}_{\mathcal{G}} \to _\} \;.$$

  As $\mathscr{A}(\mathsf{vrt})$ is deterministic, $\Delta_{\mathsf{vrt}}$ is implemented as a table with rows indexed by $\mathbb{Q}_{\mathsf{vrt}}$ and columns indexed by input symbols $\Theta$. Filling this table is done in time $\mathcal{O}(|\Theta|^2)$.

**Lemma 7.1.** *If* $\mathcal{G}$ *is complete, then* $[\![\mathscr{A}(\mathsf{vrt})]\!] = \mathcal{G}^{\uparrow}(\mathsf{vrt})$.

- For every $A \in \Gamma$ automaton $\mathscr{A}(A)$ over $\Theta$ is like $\mathsf{vrt}$ except for its final states, namely $\mathscr{A}(A) := (\mathbb{Q}_{\mathsf{vrt}}, \Delta_{\mathsf{vrt}}, \iota_{\mathsf{vrt}}, \mathbb{F}_A)$ where $\mathbb{F}_A := \{i \in \mathbb{F}_{\mathsf{vrt}} : \ell(i) = A\}$. It is built on-the-fly when constructing $\mathscr{A}(\mathsf{vrt})$. It shares its transition table with $\mathscr{A}(\mathsf{vrt})$ and needs only different marking of final states. Such marking requires linear time with respect to the number of states.

**Corollary 7.2.** *If $\mathcal{G}$ is complete, then $[\![\mathscr{A}(A)]\!] = \mathcal{G}^{\uparrow}(A)$ for every $A \in \Gamma$.*

- Let $\Sigma_\diamond := \overrightarrow{\Sigma} \cup \overleftarrow{\Sigma} \cup \{\diamond\}$ with $\diamond \notin \Sigma$ and $\Omega_\square := \Omega \cup \{\square\}$ with $\square \notin \Omega$. We define $\mathscr{A}_{\mathfrak{I}} := (\mathbb{Q}_{\mathfrak{I}}, \Delta_{\mathfrak{I}}, \iota_{\mathfrak{I}}, \mathbb{F}_{\mathfrak{I}})$, running over incidences, where
$$\begin{aligned} \mathbb{Q}_{\mathfrak{I}} &:= \Sigma_\diamond \times \mathbb{Q}_{\mathsf{vrt}} \times \Omega_\square \times \mathbb{Q}_{\mathsf{vrt}}, \\ \iota_{\mathfrak{I}} &:= (\diamond, \iota_{\mathsf{vrt}}, \square, \iota_{\mathsf{vrt}}), \\ \mathbb{F}_{\mathfrak{I}} &:= (\overrightarrow{\Sigma} \cup \overleftarrow{\Sigma}) \times \mathbb{F}_{\mathsf{vrt}} \times \Omega \times \mathbb{F}_{\mathsf{vrt}}, \end{aligned}$$
and $\Delta_{\mathfrak{I}} \subseteq \mathbb{Q}_{\mathfrak{I}} \times \Theta \times (\Theta \cup \Omega) \times \Theta \times \mathbb{Q}_{\mathfrak{I}}$ is defined assuming the following shorthands:
  - $\mathrm{prd}(i) \xrightarrow{\overleftrightarrow{b}} \mathrm{prd}(k) \to \sqsupset$ stands for $\mathrm{prd}(i) \xrightarrow{b} \mathrm{prd}(k) \to \sqsupset$ when $\overleftrightarrow{b} = \overrightarrow{b}$ and for $\mathrm{prd}(k) \xrightarrow{b} \mathrm{prd}(i) \to \sqsupset$ when $\overleftrightarrow{b} = \overleftarrow{b}$,
  - $[\overleftrightarrow{a}, i, j, k]_{\mathfrak{I}}$ means $(\overrightarrow{a}, i, j, k)$ and one has $\{a(i,j,k), \overrightarrow{a}(i,j,k), \overleftarrow{a}(k,j,i)\} \cap \sqsupset \neq \varnothing$, or it means $(\overleftarrow{a}, i, j, k)$ and one has $\{a(k,j,i), \overrightarrow{a}(k,j,i), \overleftarrow{a}(i,j,k)\} \cap \sqsupset \neq \varnothing$.

  Using these shorthands, $\Delta_{\mathfrak{I}}$ is defined as follows:
$$\begin{aligned} \Delta_{\mathfrak{I}} := & \Big\{(\overleftrightarrow{b}, i', j', k') \xrightarrow{(i,j,k)} [\overleftrightarrow{a}, i, j, k]_{\mathfrak{I}} \ : \ \mathrm{prd}(j) = \mathrm{prd}(i) \xrightarrow{\overleftrightarrow{b}} \mathrm{prd}(k) \to \sqsupset \in \mathsf{E}_{\mathcal{G}} \ \wedge \\ & \qquad \mathrm{prd}(i) = \ell(i') \to _ \in \mathsf{V}_{\mathcal{G}} \ \wedge \ \mathrm{prd}(k) = \ell(k') \to _ \in \mathsf{V}_{\mathcal{G}} \Big\} \cup \\ & \{(\diamond, i', \square, i') \xrightarrow{(i,j,k)} (\overrightarrow{a}, i, j, k) \ : \ \ell(i') \to (_,_,_,\beth) \in \mathsf{V}_{\mathcal{G}} \ \wedge \ a(i,j,k) \in \beth\} \cup \\ & \{(\diamond, i', \square, i') \xrightarrow{(i,i,i)} (\diamond, i, \square, i) \ : \ i' \xrightarrow{i} i \in \Delta_{\mathsf{vrt}}\} \cup \\ & \{\iota_{\mathfrak{I}} \xrightarrow{(i,i,i)} (\diamond, i, \square, i) \ : \ \iota_{\mathsf{vrt}} \xrightarrow{i} i \in \Delta_{\mathsf{vrt}}\} \cup \\ & \{\iota_{\mathfrak{I}} \xrightarrow{(i,j,k)} (\overrightarrow{a}, i, j, , k) \ : \ \mathsf{A}_{\mathcal{G}} \to (_,_,_,\beth) \wedge a(i,j,k) \in \beth\} \ . \end{aligned}$$

As $\mathscr{A}_{\mathfrak{I}}$ is deterministic and every accessible state $(\overleftrightarrow{c}, _, _, _) \in \mathbb{Q}_{\mathfrak{I}}$ is in $\mathbb{F}_{\mathfrak{I}}$, the following notation makes sense $(\!|\mathscr{A}_{\mathfrak{I}}|\!) := \{\overleftrightarrow{c}(u,w,v) \ : \ \iota_{\mathfrak{I}} \overset{\otimes(u,w,v)}{\dashrightarrow} (\overleftrightarrow{c}, _, _, _)\}$, with dashed arrow representing a path of $\mathscr{A}_{\mathfrak{I}}$. By considering that every $\overleftrightarrow{c} \in \overrightarrow{\Sigma} \cup \overleftarrow{\Sigma}$ represents an operation mapping triples to incidences, $\overrightarrow{c} : (u,w,v) \mapsto c(u,w,v)$ and $\overleftarrow{c} : (u,w,v) \mapsto c(v,w,u)$, the soundness of the construction of $\mathscr{A}_{\mathfrak{I}}$ is stated as follows.

**Lemma 7.3.** *If $\mathcal{G}$ is complete, then $(\!|\mathscr{A}_{\mathfrak{I}}|\!) = \mathfrak{I}(\mathcal{G})$, where $\mathfrak{I}(\mathcal{G})$ is the set of incidences of $\mathcal{G}^{\uparrow}$.*

For understanding the time complexity of the construction of $\mathscr{A}_{\mathfrak{I}}$, remember that $\mathcal{E}(\mathcal{G})$ is in bijection with $\mathfrak{I}(\mathcal{G})$. Although $\mathbb{Q}_{\mathfrak{I}} = \Sigma_\diamond \times \mathbb{Q}_{\mathsf{vrt}} \times \Omega_\square \times \mathbb{Q}_{\mathsf{vrt}}$, for every $j \in \Omega$, there is exactly one state $(_,_,j,_) \in \mathbb{Q}_{\mathfrak{I}}$. Indeed, for subset $\Sigma_\diamond \times \mathbb{Q}_{\mathsf{vrt}} \times \Omega \times \mathbb{Q}_{\mathsf{vrt}}$ of $\mathbb{Q}_{\mathfrak{I}}$, its components other than $\Omega$ aim only at keeping their track for making the definition and the proof simple. Thus, the transition table storing $\Delta_{\mathfrak{I}}$ for this subset of $\mathbb{Q}_{\mathfrak{I}}$, can be thought as having its rows indexed by $\Omega$. For subset $\Sigma_\diamond \times \mathbb{Q}_{\mathsf{vrt}} \times \{\square\} \times \mathbb{Q}_{\mathsf{vrt}}$ of $\mathbb{Q}_{\mathfrak{I}}$, $\Delta_{\mathfrak{I}}$ is isomorphic to $\Delta_{\mathsf{vrt}}$. We do not need to construct this part of $\Delta_{\mathfrak{I}}$ explicitly. We can plug $\Delta_{\mathsf{vrt}}$ into $\Delta_{\mathfrak{I}}$ via transitions corresponding to the second member of the union in the above definition of $\Delta_{\mathfrak{I}}$. When so plugging $\mathscr{A}(\mathsf{vrt})$ into the optimised version of $\mathscr{A}_{\mathfrak{I}}$, we only need to "triple" the input word of $\mathscr{A}(\mathsf{vrt})$, say $u$. We consider that $\otimes(u,u,u)$ instead $u$ is being read when switching from $\mathscr{A}(\mathsf{vrt})$ to the optimised version of $\mathscr{A}_{\mathfrak{I}}$. Thus, when constructing this version, we only need to build from scratch the portion of $\Delta_{\mathfrak{I}}$ with rows indexed by $\Omega$. Although the columns for this portion are indexed by triples in $\Theta \times \Omega \times \Theta$, only those triples that correspond to incidences in $\mathrm{rhs}(\mathsf{V}_{\mathcal{G}})$ are relevant. The number of such incidences is bound by $|\Omega|$. Consequently, the time complexity of this construction is in $\mathcal{O}(|\Omega|^2)$.

Edge related automata are obtained from auxiliary automaton $\mathscr{A}_{\mathcal{I}}$. For every $\overleftrightarrow{a} \in \overrightarrow{\Sigma} \cup \overleftarrow{\Sigma}$, we define $\mathscr{A}(\overleftrightarrow{a})$ which is like $\mathscr{A}_{\mathcal{I}}$, except for its final states $\mathbb{F}_{\overleftrightarrow{a}} := \{\overleftrightarrow{a}\} \times \mathbb{F}_{\mathsf{vrt}} \times \Omega \times \mathbb{F}_{\mathsf{vrt}}$. Then

- for every $a \in \Sigma$, we define $\mathscr{A}(a) := \mathscr{A}(\overrightarrow{a}) \cup \mathrm{swap}_{1\leftrightarrow 3}(\mathscr{A}(\overleftarrow{a}))$ where $\mathrm{swap}_{1\leftrightarrow 3}$ only modifies transitions' labels by exchanging their 1st and 3rd components:
$$\mathrm{swap}_{1\leftrightarrow 3}(\Delta_{\overleftarrow{a}}) := \{p \xrightarrow{(i,j,k)} q \ : \ p \xrightarrow{(k,j,i)} q \in \Delta_{\overleftarrow{a}}\}.$$
- $\mathscr{A}(\mathsf{edg})$ is obtained by projecting $\mathscr{A}_{\mathcal{I}}$ on the second component: $\mathscr{A}(\mathsf{edg}) = \pi_2(\mathscr{A}_{\mathcal{I}})$. Note that this projection preserves the determinism, again, thanks to the bijection between edges and incidences.

All these automata share their transition table with $\mathscr{A}_{\mathcal{I}}$ and are derived from the latter in linear time.

**Corollary 7.4.** *If $\mathcal{G}$ is complete, then $[\![\mathscr{A}(a)]\!] = \mathcal{G}^{\uparrow}(a)$ for every $a \in \Gamma$ and $[\![\mathscr{A}(\mathsf{edg})]\!] = \mathcal{E}(\mathcal{G})$.*

Since the compatibility relation comes as a bonus that is not essential for the other results, before focusing on it, we summarise the complexity of the basic construction.

**Theorem 7.5.** *For every complete graph grammar $\mathcal{G} \in 0\mathrm{L}(_,_,\Theta,\Omega)$, one can construct in time $\mathcal{O}(|\Theta|^2 + |\Omega|^2)$ an automatic presentation of $\mathcal{G}^{\uparrow}$ without the compatibility relation.*

It remains to define $\mathscr{A}(\circ)$. Although its direct construction has a better complexity, to keep this presentation simple, we get $\mathscr{A}(\circ)$ as the complement $\mathscr{A}(\between)$ which in turn is obtained as an adequate query automaton. We start with $\mathscr{A}(\between_0)$. Observe first that, as $\mathscr{A}_{\mathcal{I}}$ mimics $\mathscr{A}(\mathsf{vrt})$ before reaching its accepting states, if we modify its projection $\mathscr{A}(\mathsf{edg})$ by making accepting all its states but $\iota_{\mathcal{I}}$, we get previsely $\mathscr{A}_{\delta}$: $[\![\mathscr{A}_{\delta}]\!] = \mathcal{V}(\mathcal{G}) \cup \mathcal{E}(\mathcal{G})$. Let then $\mathscr{B}_1 := \mathscr{A}_{\delta}^{=}$ and $\mathscr{B}_2 := \mathscr{A}_{\delta}^2$ be the loose product of $\mathscr{A}_{\delta}$ by itself. Observe that $[\![\mathscr{B}_1]\!]$ is the equality on $\mathcal{V}(\mathcal{G}) \cup \mathcal{E}(\mathcal{G})$ and $[\![\mathscr{B}_2]\!] = \bigcup_{l\in\mathbb{N}}(\mathcal{V}_l(\mathcal{G}) \cup \mathcal{E}_l(\mathcal{G}))^2$. Note also that $\mathscr{B}_1$ and $\mathscr{B}_2$ have disjoint sets of states. Then, we define $\mathscr{A}(\between_0)$ as their concatenation $\mathscr{A}(\between_0) := \mathscr{B}_1 \xrightarrow{\Delta} \mathscr{B}_2$where $\Delta$ switches the control from $\mathscr{B}_1$ to $\mathscr{B}_2$ when two distinct productions apply on a same state. Assuming that $\Delta_1$ is the set of transitions of $\mathscr{B}_1$, the switch is defined as follows:
$$\Delta := \left\{p \xrightarrow{(i,j)} (q,r) \ : \ \{p \xrightarrow{i} q, p \xrightarrow{j} r\} \subseteq \Delta_1 \ \wedge \ \mathrm{lhs}(\mathrm{prd}(i)) = \mathrm{lhs}(\mathrm{prd}(j)) \ \wedge \ \mathrm{prd}(i) \neq \mathrm{prd}(j)\right\} .$$
The following lemma attests the soundness of this construction.

**Lemma 7.6.** *If $\mathcal{G}$ is complete, then $[\![\mathscr{A}(\between_0)]\!] = \between_0$.*

Observe that, $\mathscr{B}_1$ is derived from $\mathscr{A}_{\mathcal{I}}$ in linear time. However $\mathscr{B}_2$ is quadratic with respect to $\mathscr{A}_{\mathcal{I}}$. Therefore, $\mathscr{A}(\between_0)$ is built in time $\mathcal{O}(|\Theta \cup \Omega|^4)$, because the time of making $\Delta$ is in $\mathcal{O}(|\Theta \cup \Omega|^2)$.

Finally, observe that $\between$ is FO definable using relations so far defined by automata:
$$x \between y \ :\Leftrightarrow \ x \between_0 y \ \vee \ \exists\, x_1 x_2 x_3 y_1 y_2 y_3 \Big((x_1,x_2,x_3) \in \mathcal{I}(\mathcal{G}) \wedge (y_1,y_2,y_3) \in \mathcal{I}(\mathcal{G}) \wedge \Big(\bigvee_{i,j\in\{1,3\}} x_i \between_0 y_j\Big) \wedge$$
$$\Big(\Big(\Big(\bigvee_{i\in[3]} x_i \sqsubseteq x\Big) \wedge y_2 \sqsubseteq y\Big) \vee \Big(\Big(\bigvee_{i\in[3]} y_i \sqsubseteq y\Big) \wedge x_2 \sqsubseteq x\Big)\Big)\Big),$$
where $(x,y,z) \in \mathcal{I}(\mathcal{G})$ is a shorthand for $\bigvee_{a\in\Sigma} a(x,y,z)$. The corresponding query automaton $\mathscr{A}(\between)$ is built from $\mathscr{A}(\between_0)$, $\mathscr{A}_{\mathcal{I}}$ and $\mathscr{A}(\sqsubseteq)$. Then $\mathscr{A}(\circ)$ is obtained as the complement of $\mathscr{A}(\between)$.

The overall soundness of the automatic presentation of $\mathcal{G}^{\uparrow}$ developed in this section is stated in the following theorem.

**Theorem 7.7.** *For every complete 0L graph grammar $\mathcal{G}$, the result of every* FO($\Upsilon$) *query on $\mathcal{G}^{\uparrow}$ is effectively given by a finite automaton.*

Thm. 6.1 is a direct consequence of the above construction. Corollary 6.4 derives from the latter and from Thm. 6.2.

# 8 Undecidablility of language FO-checking for 0L graph languages

Unfortunately, in the unrestricted case, most of decision problems for 0L graph languages considered so far turn out to be undecidable. In this section, we construct a 0L graph grammar simulating a deterministic Turing machine. The resulting grammar is edge-deterministic and unix. All undecidability results of this paper follow either directly from the construction or from its variants.

A transition rule $pA \to \overrightarrow{q}\, B$ (resp. $pA \to \overleftarrow{q}\, B$) of a Turing machine means "in state $p$ when the symbol read is $A$, overwrite it with $B$, move the head to the next (resp. previous) cell and change the state from $p$ into $q$".

Consider a deterministic Turing machine $\mathcal{T} = (\mathbb{Q}, \Delta, \iota, f)$ on some alphabet $\Xi$ and let $\Xi_{\square} := \Xi \cup \{\square\}$. To $\mathcal{T}$, we associate eD0L graph grammar $\mathcal{G}_{\mathcal{T}}$ with vertex labels $\Gamma := \Xi_{\square} \cup \{\triangleright, \triangleleft\} \cup \{pA : p \in \mathbb{Q} \wedge A \in \Xi_{\square}\}$ and unlabelled edges in such a way that

$$\varepsilon \in \mathscr{L}(\mathcal{T}) \quad \Leftrightarrow \quad \exists G \in \mathscr{L}(\mathcal{G}_{\mathcal{T}}) \;\; G \vDash \exists x\, f(x) \ . \tag{8.1}$$

We list the elements of $\Xi_{\square}$ (resp. $\Xi_{\square,\triangleright} := \Xi_{\square} \cup \{\triangleright\}$) as $\Xi_{\square} = \{C_1, \ldots, C_n\}$ (resp. $\Xi_{\square,\triangleright} = \{D_1, \ldots, D_{n+1}\}$). Grammar $\mathcal{G}_{\mathcal{T}}$ has the following vertex productions

$$\begin{aligned} \mathsf{V}_{\mathcal{G}_{\mathcal{T}}} \;:=\; & \{\triangleleft \to (\square \to \triangleleft)\} \cup \{C \to C \;:\; C \in \Xi_{\square}\} \cup \{pA \to A \;:\; pA \to _ \in \Delta\} \cup \\ & \{C \to qC \;:\; C \in \Xi_{\square} \;\wedge\; _ \to \overleftrightarrow{q}\, _ \in \Delta\} \cup \{\mathsf{A}_{\mathcal{G}_{\mathcal{T}}} \to (\triangleright \to \iota\square \to \square \to \triangleleft)\} \end{aligned}$$

with axiom $\mathsf{A}_{\mathcal{G}_{\mathcal{T}}}$. Note that $\mathcal{G}_{\mathcal{T}}$ is not vD0L. Instead of an exhaustive listing of edge productions in $\mathsf{E}_{\mathcal{G}_{\mathcal{T}}}$, $\mathcal{G}_{\mathcal{T}}$ is given using the graphical formalism introduced in Sect. 4, except that, for better readability, vertex production are explicit and use dashed arrows in this representation. Moreover, every matching edge is drawn in blue and the inherited edge is drawn in magenta.

- The following edge production "extends" the tape of $\mathcal{T}$ to the right

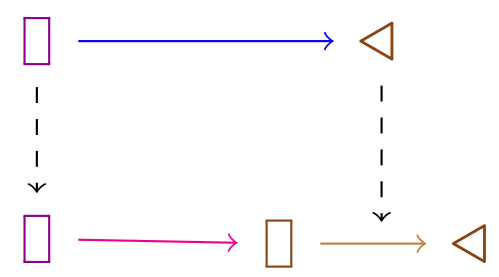

  so that the simulated head of $\mathcal{T}$ cannot reach $\triangleleft$.
- For all $D \in \Xi_{\square,\triangleright}$ and $C \in \Xi_{\square}$, we have edge productions

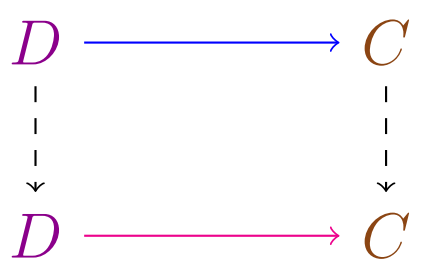

- For each rule $pA \to \overrightarrow{q}\, B \in \Delta$, we have edge productions

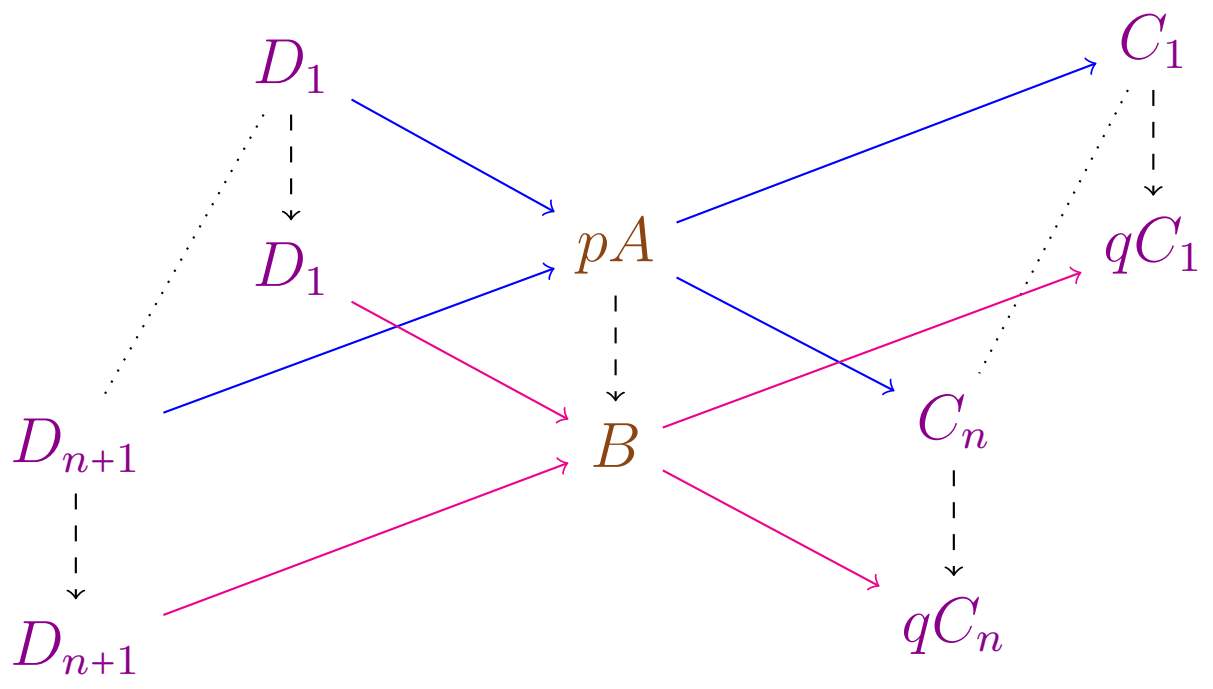

- For each rule $pA \to \overleftarrow{q}\, B \in \Delta$, we have edge productions

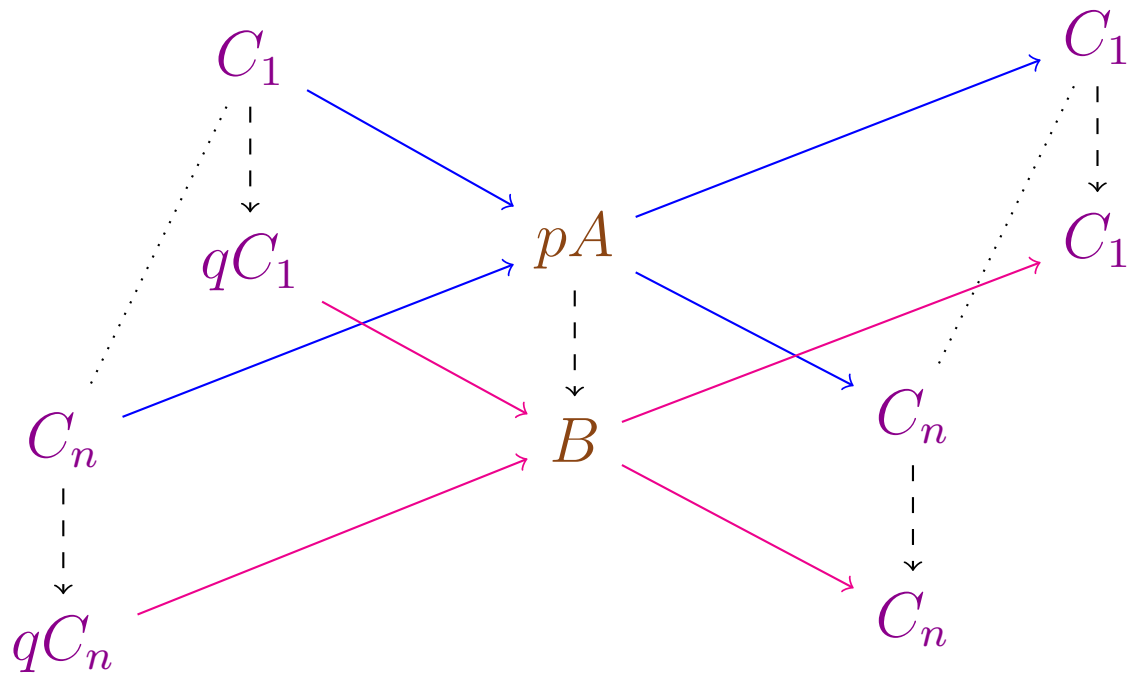

Observe that every $pA$ in the left hand side of $\Delta$ uniquely determines a vertex production of $\mathsf{V}_{\mathcal{G}_{\mathcal{T}}}$ because $\mathcal{T}$ is deterministic. Then, that vertex production falls into one and only one of the two latter diagrams of edge productions. We get eD0L graph grammar $\mathcal{G}_{\mathcal{T}}$ where every derived graph is a word with the corresponding successor represented by unlabelled edges. Moreover, each such word represents a reachable configuration of $\mathcal{T}$ possibly padded with some blanks on the right and ended by $\lhd$. Indeed, by rewriting the graph given by the axiom, there is always only one vertex with a label of the form $pA$ and all other vertices have a label in $\Xi_{\square}$ except the leftmost one labelled $\rhd$ and rightmost one labelled $\lhd$. Therefore, (8.1) holds. Note that $\mathcal{G}_{\mathcal{T}}$ is not complete. However it is unix. Consequently, we have the following:

**Proposition 8.1.**
(1) *The model* $\mathrm{FO}(\Gamma \cup \Sigma)$*-checking problem is undecidable for the class of structures*
$$\{\mathcal{G}^{\uparrow} \ : \ \mathcal{G} \text{ is eD0L and unix}\} \ .$$
(2) *The language* $\mathrm{FO}(\Gamma \cup \Sigma)$*-checking problem is undecidable for graph languages generated by unix eD0L graph grammars.*

Many other undecidability results derive from this construction. Here is an example.

**Proposition 8.2.** *The following problems*

| | |
|---|---|
| *Instance:* | *graph grammar* $\mathcal{G}$. |
| *Connectedness:* | *Are all graphs in* $\mathscr{L}(\mathcal{G})$ *(strongly) connected?* |
| *Hamiltonicity:* | *Do all graphs in* $\mathscr{L}(\mathcal{G})$ *have a Hamiltonian path?* |

*are undecidable for unix eD0L graph grammars.*

*Proof.* We extend the construction of $\mathcal{G}_{\mathcal{T}}$ by adding erasing productions for vertex labels involving the final state of $\mathcal{T}$. Then both connectedness and Hamiltonicity are equivalent to $\varepsilon \notin \mathscr{L}(\mathcal{T})$. □

A slightly different consequence of the construction of $\mathcal{G}_{\mathcal{T}}$ is stated in the following.

**Proposition 8.3.** *The language* $\mathrm{FO}(\Gamma \cup \Sigma)$*-checking problem is undecidable for graph languages generated by complete eD0L graph grammars*

*Proof.* Let $\mathcal{G}_{\mathcal{T}}^{\mathsf{i!}}$ be the completion of $\mathcal{G}_{\mathcal{T}}$ stipulated by Prop. 4.2. Observe that translation $\tau_{\mathsf{i!}}$ from the proof of Corollary 6.4, allows reducing the question "$\varepsilon \in \mathscr{L}(\mathcal{T})$ ?" via the construction of $\mathcal{G}_{\mathcal{T}}$ to the question whether there exists $G \in \mathscr{L}(\mathcal{G}_{\mathcal{T}}^{\mathsf{i!}})_{\{\mathsf{i},\mathsf{!}\}}$ such that $G \models \tau_{\mathsf{i!}}(\exists\, x\, f(x))$. □

From this proposition and Thm. 6.1, we conclude that the decidability of $\mathrm{FO}(\Gamma \cup \Sigma)$-theory of complete 0L grammars is not sufficient for the decidability of the $\mathrm{FO}(\Gamma \cup \Sigma)$-language checking problem for complete eD0L graph grammars. In the presence of vertex nondeterminism, it is not possible in general to "separate" graphs of the same layer sharing some vertices.

# 9 Conclusion

Although this work only addresses first-order logic, its application potential beyond mathematics and theoretical computer science would benefit from extending our approach to quantitative logics. Such extensions also seem desirable for automatic structures. To our knowledge, examples of application of automatic structures outside of mathematics are still awaited. We hope that the connection of this field of scientific achievements with expanders and Lindenmayer graph grammars proposed in this paper can stimulate even more relevant theoretical developments that would enable such applications.

# Appendices

The statements without proof in the main part appear with their original numbers in the following appendices.

## A Fundamental facts on complete and incomplete 0L graph grammars

In this appendix, the reader will find complete proofs of statements of Sect. 4.

**Lemma 4.1.** *Let $\mathcal{G} = (\mathsf{A}_\mathcal{G}, \mathsf{V}_\mathcal{G}, \mathsf{E}_\mathcal{G})$ be a complete graph grammar in $0\mathrm{L}(\Gamma, \Sigma, _, _)$ and let $G = (V, E, \ell, \gimel)$ be a graph in $\mathscr{L}(\mathcal{G})$. Every labelling preserving map $h{:}\, V \to \mathsf{V}_\mathcal{G}$ extends into a homomorphism from $G$ into $\mathcal{G}$.*

*Proof.* Let $h{:}\, V \to \mathsf{V}_\mathcal{G}$ be such that $h(v) = \ell(v) \to _$ for every $v \in V$. In order to extend $h$ into a homomorphism from $G$ to $\mathcal{G}$, consider $w \in E$ and the corresponding $a(u, w, v) \in \gimel$. Set $h(w) := P$ for some $P = h(u) \xrightarrow{a} h(v) \to _$. Observe that $P \in \mathsf{E}_\mathcal{G}$ because $\mathcal{G}$ is complete. Clearly, $h$ so extended onto edges is a homomorphism from $G$ to $\mathcal{G}$. □

**Proposition 4.2.** *For every 0L graph grammar $\mathcal{G} = (\mathsf{V}_\mathcal{G}, _, _)$ one can construct in time $\mathcal{O}(|\mathsf{V}_\mathcal{G}|^2)$ using additional non-terminal vertex label “ ¡ ” and non-terminal edge label “ ! ” complete 0L graph grammar $\mathcal{G}^{¡!}$ such that $\mathscr{L}(\mathcal{G}) = \mathscr{L}(\mathcal{G}^{¡!})_{\{¡,!\}}$. If $\mathcal{G}$ is vD0L or eD0L, then $\mathcal{G}^{¡!}$ is so.*

*Proof.* Let $\mathcal{G} = (\mathsf{V}_\mathcal{G}, \mathsf{E}_\mathcal{G}, \mathsf{A}_\mathcal{G})$ be a 0L graph grammar in $0\mathrm{L}(\Gamma, \Sigma, \Theta, \Omega)$. Consider bijection $g_1{:}\, \Theta' \to \Gamma' \cup \{¡\}$ where $\Theta'$ is a new set of vertex names $\Theta' \cap \Theta = \varnothing$, $¡ \notin \Gamma$ is a new vertex label and $\Gamma'$ consists of labels missing in $\mathrm{lhs}(\mathsf{V}_\mathcal{G})$ but labelling isolated vertices in $\mathrm{rhs}(\mathsf{V}_\mathcal{G})$. Let then $\mathcal{G}^{¡} := (\mathsf{V}_\mathcal{G} \cup \mathsf{V}^{¡}, \mathsf{E}_\mathcal{G}, \mathsf{A}_\mathcal{G})$ where $\mathsf{V}^{¡} := \big\{ g_1(i) \to (\{i\}, \varnothing, \{i \mapsto ¡\}, \varnothing) : i \in \Theta' \big\}$. Every right hand side of $\mathsf{V}^{¡}$ is a graph with a single ¡-labelled vertex with no loop. Observe that a possible incompleteness of $\mathcal{G}^{¡}$ can only be due to edge productions but no more to vertex productions. Observe also that if $\mathcal{G}$ is vD0L or eD0L, then $\mathcal{G}^{¡}$ is so.

Next, we transform $\mathcal{G}^{¡}$ into the target grammar $\mathcal{G}^{¡!}$ by completing $\mathsf{E}_{\mathcal{G}}$. Let $M$ be the set of all triples in $\mathsf{V}_{\mathcal{G}} \times \Sigma \times \mathsf{V}_{\mathcal{G}}$ that are not among left hand sides of $\mathsf{E}_{\mathcal{G}}$ but which could be matched by incidences appearing in right hand sides of productions of $\mathcal{G}$. In symbols

$$M := \{P_1 \xrightarrow{a} P_2 \in \mathsf{V}_{\mathcal{G}} \times \Sigma \times \mathsf{V}_{\mathcal{G}} \setminus \mathrm{lhs}(\mathsf{E}_{\mathcal{G}}) \ : \ \exists\, a(i,k,j) \in \sqsupset_{\mathcal{G}} \ \ell_{\mathcal{G}}(i) = \mathrm{lhs}(P_1) \wedge \ell_{\mathcal{G}}(j) = (\mathrm{lhs}(P_2)\}$$

where $\sqsupset_{\mathcal{G}} := \bigcup \pi_4(\mathrm{rhs}(\mathsf{V}_{\mathcal{G}})) \cup \bigcup \mathrm{rhs}(\mathsf{E}_{\mathcal{G}})$ and $\ell_{\mathcal{G}} := \bigcup \pi_3(\mathrm{rhs}(\mathsf{V}_{\mathcal{G}}))$. Consider bijection $g_2 \colon \Omega' \to M$ for some set of new edge names $\Omega'$ such that $\Omega' \cap (\Theta \cup \Theta' \cup \Omega) = \varnothing$ and injective mapping $f \colon \mathsf{V}_{\mathcal{G}} \to \Theta$ such that $f(A \to (V, _, _, _)) \in V$ for every $A \to (V, _, _, _) \in \mathsf{V}_{\mathcal{G}}$. Set

$$\mathsf{E} := \left\{P_1 \xrightarrow{a} P_2 \to \{!(f(P_1), i, f(P_2))\} \ : \ i \in \Omega' \wedge g_2(i) = P_1 \xrightarrow{a} P_2\right\} \ .$$

As every triple in $M$ witnesses the incompleteness of $\mathcal{G}^{¡}$, $\mathsf{E}$ form a partial completion of $\mathcal{G}^{¡}$. Indeed, all incidences of $\sqsupset_{\mathcal{G}}$ have a match in $\mathsf{E}_{\mathcal{G}} \cup \mathsf{E}$. However, adding $\mathsf{E}$ alone to $\mathsf{E}_{\mathcal{G}}$ is not enough because now we have new incidences with label ! that have no matching edge productions in $\mathsf{E}_{\mathcal{G}} \cup \mathsf{E}$. Consider therefore bijection $g_2^! \colon \Omega^! \to \mathsf{V}_{\mathcal{G}} \times \mathsf{V}_{\mathcal{G}}$ for some yet newer set of edge names $\Omega^!$ such that $\Omega^! \cap (\Theta \cup \Theta' \cup \Omega \cup \Omega') = \varnothing$ and set

$$\mathsf{E}^! := \left\{P_1 \xrightarrow{!} P_2 \to \{!(f(P_1), i, f(P_2))\} \ : \ i \in \Omega' \wedge g_2^!(i) = (P_1, P_2)\right\} \ .$$

Let then
$$\mathcal{G}^{¡!} := (\mathsf{V}_{\mathcal{G}} \cup \mathsf{V}^{¡}, \mathsf{E}_{\mathcal{G}} \cup \mathsf{E} \cup \mathsf{E}^!, \mathsf{A}_{\mathcal{G}}) \ .$$

Observe that $\mathcal{G}^{¡!} \in 0\mathrm{L}(\Gamma \cup \{¡\}, \Sigma \cup \{!\}, \Theta \cup \Theta', \Omega \cup \Omega' \cup \Omega^!)$ is complete and that $\mathscr{L}(\mathcal{G}) = \mathscr{L}(\mathcal{G}^{¡!})_{\{¡,!\}}$. Note also that $\mathcal{G}^{¡!}$ has the same vertex production as $\mathcal{G}^{¡}$. Thus, if $\mathcal{G}^{¡}$ is vD0L, $\mathcal{G}^{¡!}$ is so. To see that the transformation from $\mathcal{G}^{¡}$ to $\mathcal{G}^{¡!}$ preserves the property of being eD0L, observe first that $\mathrm{lhs}(\mathsf{E} \cup \mathsf{E}^!) \cap \mathrm{lhs}(\mathsf{E}_{\mathcal{G}}) = \varnothing$. Obviously also $\mathrm{lhs}(\mathsf{E}) \cap \mathrm{lhs}(\mathsf{E}^!) = \varnothing$. Moreover, two distinct productions of $\mathsf{E} \cup \mathsf{E}^!$ never share their left hand sides as both $g_2$ and $g_2^!$ are bijections. Consequently, $\mathcal{G}^{¡!}$ is eD0L when $\mathcal{G}_1$ is so. Hence, the transformation from $\mathcal{G}$ is both vD0L and eD0L preserving. □

**Theorem 4.9.** *The following problems*

| | |
|---|---|
| *Instance:* | *0L graph grammar $\mathcal{G}$, label $\lambda$ of $\mathcal{G}$ and production $P$ of $\mathcal{G}$.* |
| *Finiteness:* | *Is $\mathscr{L}(\mathcal{G})$ finite ?* |
| *Emptiness:* | *$\mathscr{L}(\mathcal{G}) = \varnothing$ ?* |
| *Accessibility:* | *Is $\lambda$ accessible in $\mathcal{G}$ ?* |
| *Usefulness:* | *Is $P$ useful for $\mathcal{G}$ ?* |

*are*

(i) *decidable for complete 0L graph grammars without non-terminals and for D0L graph grammars even with non-terminals,*

(ii) *undecidable for unix eD0L graph grammars, with non-terminals in the case of emptiness.*

*Proof.*

(i) Except for the emptiness, the decidability of each problem results from Thm. 6.1 by reduction to the model checking of $\mathcal{G}^{\uparrow}$ or $\mathcal{G}^{\uparrow \setminus N}$.

**Accessibility** For complete 0L graph grammar $\mathcal{G}$ and $\lambda \in \Gamma$ (resp. $\lambda \in \Sigma$) we check if $\mathcal{G}^{\uparrow}$ models $\exists x\, \lambda(x)$ (resp. $\exists xyz\, \lambda(x,y,z)$). For D0L graph grammar $\mathcal{G}$ with non-terminals $N$ we similarly check $\mathcal{G}^{\uparrow \setminus N}$.

**Usefulness** We modify $P$ by introducing an extra label to $\mathrm{rhs}(P)$ and ask if the new label is accessible.

**Finiteness** Depending on whether $\mathcal{G}$ is complete or deterministic with non-terminals $N$, we check $\mathcal{G}^{\uparrow}$ or $\mathcal{G}^{\uparrow \setminus N}$ against the negation of the following formula

$$\forall x\, \exists y\, \left(x \mathrel{\hat{=}} y \Rightarrow \exists z\, (y \neq z \wedge y \sqsubseteq z)\right) \ .$$

The formula says that every layer has an element which is a strict prefix of another element. Hence, the formula requires an infinite number of layers to hold. This is equivalent to having $\mathscr{L}(\mathcal{G})$ (resp. $\mathscr{L}(\mathcal{G})_N$) infinite when $\mathcal{G}$ is without non-terminals (resp. with non-terminals $N$).

**Emptiness** For 0L graph grammar $\mathcal{G} = (\mathsf{V}_{\mathcal{G}}, \mathsf{E}_{\mathcal{G}}, \mathsf{A}_{\mathcal{G}})$ without non-terminals, we have $\mathscr{L}(\mathcal{G}) = \varnothing$ if and only if $\mathsf{A}_{\mathcal{G}} \notin \mathrm{lhs}(\mathsf{V}_{\mathcal{G}})$. If $\mathcal{G}$ is D0L with non-terminals $N$, the emptiness is equivalent to language checking $\mathscr{L}(\mathcal{G})_N$ against a tautology, for instance $\forall x\; x{=}x$. The decidability follows from Corollary 6.4.

(ii) The undecidability of each problem results from the construction, given in Sect. 8, of unix eD0L grammar $\mathcal{G}_{\mathcal{T}}$ associated to arbitrary deterministic Turing machine $\mathcal{T}$ running on an empty tape, viz., with input word $\varepsilon$.

**Accessibility** We ask, for each vertex label $fC$ formed with accepting state $f$ of $\mathcal{T}$, if it is accessible in $\mathcal{G}_{\mathcal{T}}$. The disjunction of all answers is equivalent to $\varepsilon \in \mathscr{L}(\mathcal{T})$.

**Usefulness** We ask, for each vertex production introducing $f$, if it is useful. The disjunction of all answers is equivalent to $\varepsilon \in \mathscr{L}(\mathcal{T})$.

**Finiteness** As $\mathcal{G}_{\mathcal{T}}$ has a unique derivation, the halting of $\mathcal{T}$ is equivalent to the finiteness of $\mathscr{L}(\mathcal{G}_{\mathcal{T}})$.

**Emptiness** This problem is trivially decidable for 0L graph grammars without non-terminals (see the decidability of the emptiness above). To establish the undecidability of the emptiness for unix eD0L graph grammars with non-terminals, the construction of Sect. 8 is adapted as follows. We consider all vertex and edge labels of $\mathcal{G}_{\mathcal{T}}$ as non-terminals, say $N$, and we add one terminal vertex label $\blacklozenge$ together with vertex productions $\{fC \to \blacklozenge \;:\; C \in \Xi_{\square}\} \cup \{C \to \blacklozenge \;:\; C \in \Xi_{\triangleright,\square,\triangleleft}\}$. Moreover

- for all $D \in \Xi_{\square,\triangleright}$ and $C \in \Xi_{\square,\triangleleft}$, we add edge productions

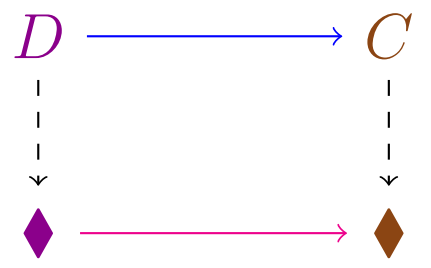

- for all $C \in \Xi_{\square}$ and $D \in \Xi_{\square,\triangleleft}$, we add edge productions

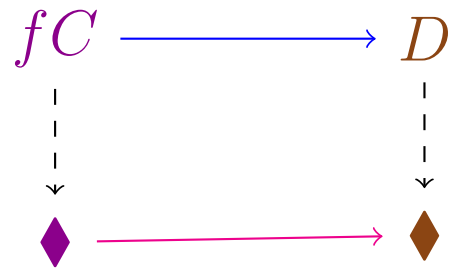

- and for all $C \in \Xi_{\square,\triangleright}$ and $D \in \Xi_{\square}$, we add edge productions

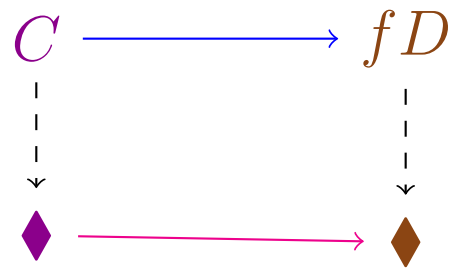

Let $\mathcal{G}'_{\mathcal{T}}$ be the grammar so extended. Observe that (see Sect. 8) reaching $f$ by $\mathcal{T}$ is reflected on $\mathcal{G}_{\mathcal{T}}$ by the unique derivation ending with a graph representing an accepting configuration with exactly one vertex labelled $fC$ for some $C \in \Xi_{\square}$. For $\mathcal{G}'_{\mathcal{T}}$, the presence of such vertex triggers the simultaneous application of new productions. The resulting graph $\blacklozenge \to \cdots \to \blacklozenge$ has the terminal label everywhere and is the unique graph in $\mathscr{L}(\mathcal{G}'_{\mathcal{T}})_N$. If, however, $\mathcal{T}$ cannot reach an accepting configuration, then $\mathscr{L}(\mathcal{G}'_{\mathcal{T}})_N = \varnothing$.

□

**Proposition 4.10.** *The (abstract) membership problem is decidable for D0L graph grammars even with non-terminals. It is undecidable for unix eD0L graph grammars.*

*Proof.* For the undecidability claim we modify the adaptation of the construction of Sect. 8 considered in the proof of Thm. 4.9 for establishing the undecidability of the emptiness problem.

The starting step of the present modification is that all labels have now the same status without distinguishing non-terminals. We take two additional vertex labels ▶ and ◀ and replace vertex productions $\triangleright \to$ ⧫ and $\triangleleft \to$ ⧫ by $\triangleright \to$ ▶ and $\triangleleft \to$ ◀ in the above adaptation. Edge productions are modified accordingly, so that we can derive graph ▶ → ⧫ → ⋯ → ⧫ → ◀ (instead of ⧫ → ⋯ → ⧫ of the above adaptation), if, and only if, $\varepsilon \in \mathscr{L}(\mathcal{T})$. We further extend the construction by adding two vertex productions ▶ → ▶, ◀ → ◀, one erasing vertex production ⧫ → ∅ and three erasing edge productions

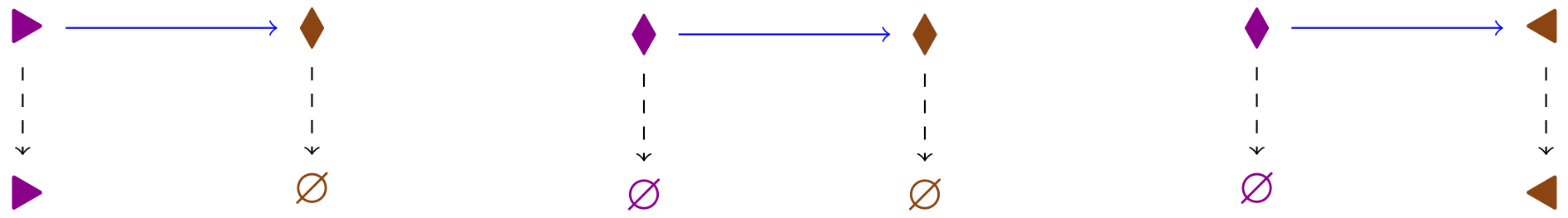

These let derive in one step from ▶ → ⧫ → ⋯ → ⧫ → ◀ the graph which consists of two isolated vertices only: ▶ ◀. The membership of this graph in the language of the modified graph grammar, which still remains unix and eD0L, is equivalent to the membership of $\varepsilon$ in $\mathscr{L}(\mathcal{T})$.

The decidability claim follows from the fact that, for every finite graph $H$, we can write a sentence $\varphi_H$ such that $\{G : G \vDash \varphi_H\}$ is the isomorphism class of $H$. Consequently, $H$ belongs up to isomorphism to $\mathscr{L}(\mathcal{G})_N$, if language checking of $\mathscr{L}(\mathcal{G})_N$ against $\varphi_H$ answers yes. This is decidable by Corollary 6.4. □

# B When graphs' elements form maximal compatible subsets of $\mathcal{V}(\mathcal{G}) \cup \mathcal{E}(\mathcal{G})$

This appendix is devoted to the proof of the the following lemma.

**Lemma 4.11.** *If $\mathcal{G}$ is a complete 0L graph grammar, then*

$$\{S \subseteq \mathcal{V}_l(\mathcal{G}) \cup \mathcal{E}_l(\mathcal{G}) : S \text{ is maximal compatible}\} = \{V \cup E : (V, E, _, _) \in \mathscr{L}_l(\mathcal{G})\}.$$

With the help of subsequent lemmas, the proof goes as follows.

*Proof.* Let $\mathcal{G}$ be a complete 0L graph grammar.

($\subseteq$) Let $S \subseteq \mathcal{V}_l(\mathcal{G}) \cup \mathcal{E}_l(\mathcal{G})$ be a maximal compatible set. By Lemma B.2, there exists $(V, E, _, _) \in \mathscr{L}_l(\mathcal{G})$ such that $S \subseteq V \cup E$. By definition of derivation, $(V \cup E)^2 \subseteq \asymp$. In other words, it is a compatible set, say $S' = V \cup E$. But as $S$ is maximal compatible, whereas $S \subseteq S'$, we have $S = S' = V \cup E$.

($\supseteq$) Let $(V, E, \ell, \daleth) \in \mathscr{L}_l(\mathcal{G})$. As observed above, $V \cup E$ is a compatible set. To see that it is maximal compatible, consider a compatible set $S \subseteq \mathcal{V}_l(\mathcal{G}) \cup \mathcal{E}_l(\mathcal{G})$ such that $V \cup E \subseteq S$. By Lemma B.2, there exists $(V', E', \ell', \daleth') \in \mathscr{L}_l(\mathcal{G})$ such that $S \subseteq V' \cup E'$. Consequently $V \cup E \subseteq V' \cup E'$. As $\Theta \cap \Omega = \varnothing$, the latter inclusion splits into $V \subseteq V'$ and $E \subseteq E'$. Observe also that for a vertex (resp. edge), say $si$ in $\mathcal{V}_l(\mathcal{G})$ (resp. in $\mathcal{E}_l(\mathcal{G})$), where $i \in \Theta$ (resp. $i \in \Omega$), its label is uniquely determined within the corresponding production $\mathrm{prd}(i)$. Moreover, when $si$ is an edge, $\mathrm{prd}(i)$ also uniquely determines the incidence of $si$. Consequently, from inclusions $V \subseteq V'$ and $E \subseteq E'$, we obtain $\ell \subseteq \ell'$ and $\daleth \subseteq \daleth'$. As $(V, E, \ell, \daleth) \subseteq (V', E', \ell', \daleth')$, by Lemma B.1, $(V, E, \ell, \daleth) = (V', E', \ell', \daleth')$. Hence, $S \subseteq (V, E, _, _)$ and we conclude that $V \cup E$ is maximal compatible. □

**Lemma B.1.** *For every 0L graph grammar $\mathcal{G}$ and all $G, G' \in \mathscr{L}_l(\mathcal{G})$, if $G \subseteq G'$ then $G = G'$.*

*Proof.* By induction on $l \in \mathbb{N}_{>0}$. Let $\mathcal{G} = (\mathsf{A}_\mathcal{G}, \mathsf{V}_\mathcal{G}, \mathsf{E}_\mathcal{G})$ be a 0L graph grammar and let $G_1 = (V_1, E_1, _, _) \in \mathscr{L}_l(\mathcal{G})$ and $G_2 = (V_2, E_2, _, _) \in \mathscr{L}_l(\mathcal{G})$ be such that $V_1 \subseteq V_2$ and $E_1 \subseteq E_2$.

($\boldsymbol{l = 0}$) Assume by contradiction that $G_1 \neq G_2$. Then $\mathcal{G}$ has productions $\mathsf{A}_\mathcal{G} \to G_1$ and $\mathsf{A}_\mathcal{G} \to G_2$ but the definition of 0L graph grammar requires that $(V_1 \cup E_1) \cap (V_2 \cup E_2) = \varnothing$ which contradicts $V_1 \subseteq V_2$ and $E_1 \subseteq E_2$.

($\boldsymbol{l > 0}$) By definition of derivation, there exist $G'_1, G'_2 \in \mathscr{L}_{l-1}(\mathcal{G})$ and homomorphisms $h_1{:}\, G'_1 \to \mathcal{G}$, $h_2{:}\, G'_2 \to \mathcal{G}$ such that $G'_1 \xrightarrow[\mathcal{G}]{h_1} G_1$ and $G'_2 \xrightarrow[\mathcal{G}]{h_2} G_2$. As $G_1 \subseteq G_2$, also $G'_1 \subseteq G'_2$. By induction hypothesis $G'_1 = (V', E', _, _) = G'_2$ for some $V' \subseteq \mathcal{V}_{l-1}(\mathcal{G})$ and $E' \subseteq \mathcal{E}_{l-1}(\mathcal{G})$. To show that $h_1 = h_2$, assume by contradiction that there exists $w \in V' \cup E'$ such that $h_1(w) = P_1 \neq P_2 = h_2(w)$ for some productions $P_1, P_2 \in \mathsf{V}_\mathcal{G} \cup \mathsf{E}_\mathcal{G}$. By definition of 0L graph grammar, the set of vertex or edge names appearing in rhs$(P_1)$ is disjoint from that appearing in rhs$(P_2)$. Then also $\{s \in V_1 \cup E_1 : w \xrightarrow[\mathcal{G}]{P_1} s\} \cap \{s \in V_2 \cup E_2 : w \xrightarrow[\mathcal{G}]{P_2} s\} = \varnothing$ which contradicts $G_1 \subseteq G_2$. □

**Lemma B.2.** *Let $\mathcal{G} \in 0\mathrm{L}(\Gamma, \Sigma, \Theta, \Omega)$ be complete and $S \subseteq \mathcal{V}_l(\mathcal{G}) \cup \mathcal{E}_l(\mathcal{G})$ be such that $S^2 \subseteq \text{ↄ}$. Then $S \subseteq V \cup E$ for some $(V, E, _, _) \in \mathscr{L}_l(\mathcal{G})$.*

*Proof.* By induction on product order $(\mathbb{N}, \leq)^2$ over pairs $(l, m)$ with $m = |S|$. Let $\mathcal{G} = (\mathsf{V}_\mathcal{G}, \mathsf{E}_\mathcal{G}, \mathsf{A}_\mathcal{G})$.

($\boldsymbol{l = 1}$) Then $S \subseteq V \cup E$ for some production $\mathsf{A}_\mathcal{G} \to (V, E, _, _) \in \mathsf{V}_\mathcal{G}$.

($\boldsymbol{m = 0}$) Trivial case.

($\boldsymbol{l > 1}$) $S$ can be written as $S = \{s_j i_j : j \in [m]\}$ with $S' = \{s_j : j \in [m]\} \subseteq \mathcal{V}_{l-1}(\mathcal{G}) \cup \mathcal{E}_{l-1}(\mathcal{G})$ and $\{i_j : j \in [m]\} \subseteq \Theta \cup \Omega$. Let $si \in S$ with $s \in S'$ and let $S_1 := S \setminus \{si\}$. As $si \in \mathcal{V}_l(\mathcal{G}) \cup \mathcal{E}_l(\mathcal{G})$, there exists a graph, say $G_2 = (V_2, E_2, _, _) \in \mathscr{L}_l(\mathcal{G})$, such that $si \in V_2 \cup E_2$. By definition of ↄ we have $S_1^2 \subseteq \text{ↄ}$ and as $|S_1| < m$, by induction hypothesis, there exists $G_1 = (V_1, E_1, _, \gimel_1) \in \mathscr{L}_l(\mathcal{G})$ such that $S_1 \subseteq V_1 \cup E_1$. Observe that $(S')^2 \subseteq \text{ↄ}$, by definition of ↄ. As also $S' \subseteq \mathcal{V}_{l-1}(\mathcal{G}) \cup \mathcal{E}_{l-1}(\mathcal{G})$, by induction hypothesis, there exists $G' = (V', E', _, \gimel') \in \mathscr{L}_{l-1}(\mathcal{G})$ such that $S' \subseteq V' \cup E'$. By definition of derivation, there exist homomorphisms $h_1, h_2{:}\, G' \to \mathcal{G}$ such that $G' \xrightarrow{h_1} G_1$ and $G' \xrightarrow{h_2} G_2$. If $h_2(s) = h_1(s)$, then $si \in V_1 \cup E_1$ and we are done because $S \subseteq V_1 \cup E_1$. Assume therefore that $h_2(s) \neq h_1(s)$. For any $t \in V'$, we denote by $E'_t$ the set of edges of $E'$ incident to $t$:

$$E'_t := \left\{ w \in E' \ : \ \{_(t, w, _), _(_, w, t)\} \cap \gimel' \neq \varnothing \right\} .$$

Using this notation, we construct homomorphism $h{:}\, G' \to \mathcal{G}$ setting $h(s) := h_2(s)$ and such that $G' \xrightarrow{h} G$ for some $G = (V, E, _, _) \in \mathscr{L}_l(\mathcal{G})$ satisfying $S \subseteq V \cup E$. The construction has two cases.

- ($\boldsymbol{s \in V'}$) Observe that $E'_s \cap S' = \varnothing$. If instead $s_k \in E'_s$ for some $k \in [m]$, then we would have $\{_(s, s_k, _), _(_, s_k, s)\} \cap \gimel' \neq \varnothing$ and $\{_(si', s_k i_k, _), _(_, s_k i_k, si')\} \cap \gimel_1 \neq \varnothing$ for some vertex name $i'$ such that $\mathrm{prd}(i') = h_1(s)$. We would then conclude that $si' \mathrel{)\!(_0} si$ because $h_1(s) \neq h_2(s)$ whereas $\mathrm{lhs}(h_1(s)) = \mathrm{lhs}(h_2(s))$, and consequently we would get $si \mathrel{)\!(} s_k i_k$ since $si'$ is an extremity of $s_k i_k$ in $G_1$. As both $si$ and $s_k i_k$ are in $S$, the latter would contradict $S^2 \subseteq \text{ↄ}$. Now, as $E'_s \cap S' = \varnothing$, for every loop $w' \in E'_s$ with corresponding incidence $a(s, w', s)$, we pick $P' \in \mathsf{E}_\mathcal{G}$ such that $\mathrm{lhs}(P') = h_2(s) \xrightarrow{a} h_2(s)$ and we set $h(w') := P'$. For every edge $w \in E'_s$, with corresponding incidence $a(s, w, v) \in \gimel'$ (resp. $a(v, w, s) \in \gimel'$) and $s \neq v$, we pick $P \in \mathsf{E}_\mathcal{G}$ such that $\mathrm{lhs}(P) = h_2(s) \xrightarrow{a} h_1(v)$ (resp. $\mathrm{lhs}(P) = h_1(v) \xrightarrow{a} h_2(s)$) and we set $h(w) := P$. The existence of picks $P$ and $P'$ results from the completeness of $\mathcal{G}$. Finally, for every $t \in (V' \setminus \{s\}) \cup (E' \setminus E'_s)$, we set $h(t) := h_1(t)$. Hence $G' \xrightarrow{h} G$ for some $G = (V, E, _, _) \in \mathscr{L}_l(\mathcal{G})$ such that $S \subseteq V \cup E$.
- ($\boldsymbol{s \in E'}$) Let $a(u, s, v) \in \gimel'$ be the incidence of $s$. Observe that if there were $s_k \in (\{u, v\} \cup E'_u \cup E'_v) \cap (S' \setminus \{s\})$ for some $k \in [m]$ then we would have $si \mathrel{)\!(} s_k i_k$ (by a reasoning similar to the above case of $s \in V'$) contradicting $S \subseteq \text{ↄ}$. As $(\{u, v\} \cup E'_u \cup E'_v) \cap S' = \{s\}$, for every edge $w \in E'_u \setminus \{s\}$ with corresponding incidence $b(u,w,t) \in \beth'$ (resp. $b(t,w,u) \in \beth'$), we pick $P \in \mathsf{E}_{\mathcal{G}}$ with $\mathrm{lhs}(P) = h_2(u) \xrightarrow{b} h_1(t)$ (resp. $\mathrm{lhs}(P) = h_1(t) \xrightarrow{b} h_2(u)$) and we set $h(w) := P$. Similarly, for every edge $w \in E'_v \setminus \{s\}$ with corresponding incidence $c(v,w,t) \in \beth'$ (resp. $c(t,w,v) \in \beth'$), we pick $P \in \mathsf{E}_{\mathcal{G}}$ with $\mathrm{lhs}(P) = h_2(v) \xrightarrow{c} h_1(t)$ (resp. $\mathrm{lhs}(P) = h_1(t) \xrightarrow{c} h_2(v)$) and we set $h(w) := P$. Thanks to the completeness of $\mathcal{G}$, all those picks are possible. To close the construction, we set $h(u) := h_2(u)$, $h(v) := h_2(v)$ and, for every $t \in (V' \setminus \{u,v\}) \cup \big(E' \setminus (E'_u \cup E'_v)\big)$, we set $h(t) := h_1(t)$. □

# C Theorems of Section 6

**Theorem 6.1.** *$\mathcal{G}^{\uparrow}$ (resp. $\mathcal{G}^{\uparrow\setminus N}$) has a decidable* FO(Υ) *theory for every 0L graph grammar $\mathcal{G}$ which is complete (resp. deterministic with non-terminals $N$).*

*Proof.* For a complete 0L graph grammar, the result follows directly from Thm. 7.7. Let then $\mathcal{G}$ be a D0L graph grammar with non-terminals $N$ and consider its completion $\mathcal{G}^{¡!}$ according to Prop. 4.2. It is easy to see that $\mathcal{G}^{\uparrow\setminus N}$ can be interpreted within $(\mathcal{G}^{¡!})^{\uparrow}$ by means of an FO(Υ)-interpretation. Then, the result is obtained using the standard backwards translation associated to the interpretation. The interpretation leaves unchanged atomic predicates other than $\mathsf{vrt}$ and $\mathsf{edg}$. Formula $\theta_{\mathsf{vrt}}(x)$ selecting $\mathcal{V}(\mathcal{G})_N$ within $(\mathcal{G}^{¡!})^{\uparrow}$ brings together all vertices from layers where no labels from $\Gamma' := \{¡\} \cup (N \cap \Gamma)$ nor $\Sigma' := \{!\} \cup (N \cap \Sigma)$ appear

$$\theta_{\mathsf{vrt}}(x) \quad :\Leftrightarrow \quad \mathsf{vrt}(x) \ \land\ \forall y \ \Big(x \mathrel{\hat{=}} y \Rightarrow \neg\big(\bigvee_{\lambda\in\Gamma'} \lambda(y) \lor \exists x z \bigvee_{\lambda\in\Sigma'} \lambda(x,y,z)\big)\Big) \ .$$

Formula $\theta_{\mathsf{edg}}(x)$ selecting $\mathcal{E}(\mathcal{G})_N$ within $(\mathcal{G}^{¡!})^{\uparrow}$ is similar. Finally, formula $\delta(x)$ selecting the interpretation domain is obvious: $\delta(x) \ :\Leftrightarrow\ \theta_{\mathsf{vrt}}(x) \lor \theta_{\mathsf{edg}}(x)$. □

**Theorem 6.2.** *There is a translation $\tau$: FO(Γ ∪ Σ) → FO(Γ ∪ Σ ∪ {≙}) in time $\mathcal{O}(nk+m)$, where $m$ is the length of the input formula, $n$ is its alternation rank and $k$ is the number of its variables, such that for every unix 0L graph grammar $\mathcal{G} \in 0\mathrm{L}(\Gamma,\Sigma,_,_)$ and every formula $\varphi(\mathcal{X}) \in \mathrm{FO}(\Gamma\cup\Sigma)$, one has*

$$\mathcal{G}^{\uparrow}(\tau(\varphi(\mathcal{X})) = \bigcup_{G\in\mathscr{L}(\mathcal{G})} G(\varphi(\mathcal{X})) \ .$$

*In particular, when $\mathcal{X} = \varnothing$, there exists $G \in \mathscr{L}(\mathcal{G})$ such that $G \vDash \varphi$, if, and only if, $\mathcal{G}^{\uparrow} \vDash \tau(\varphi)$.*

*Proof.* The input formula is first put into prenex form: $Q_1\mathcal{X}_1 \ldots Q_n\mathcal{X}_n\,\psi(\mathcal{Y})$ where $\psi(\mathcal{Y})$ is quantifier-free and every $Q_i\mathcal{X}_i$ with $Q_i \in \{\exists,\forall\}$, $\mathcal{X}_i = \{x_{i1}\ldots x_{ik_i}\}$ and $Q_i \neq Q_{i+1}$ is a quantifier block replacing $Q_i\,x_{i1}\,\ldots, Q_i\,x_{ik_i}$. Next, every universal block $\forall\,\mathcal{X}_j$ is replaced with $\neg\exists\,\mathcal{X}_j\neg$. For this transformation, we need first to parse the input formula. This is done in linear time as the set of formulae is deterministic context-free. Once parsed, putting it into prenex form is done in linear time too.

For any $\mathcal{Z} = \{z_1,\ldots,z_{|\mathcal{Z}|}\}$, we set $\mathcal{Z}^{\hat{=}} := z_1 \mathrel{\hat{=}} z_2 \mathrel{\hat{=}} \cdots \mathrel{\hat{=}} z_{|\mathcal{Z}|}$. With this notation $\tau$ is defined inductively as follows for every formula $\varphi(\mathcal{X}) \in \mathrm{FO}(\Gamma\cup\Sigma)$ in prenex form:

$$\begin{aligned}
\tau(\varphi(\mathcal{X})) &= \mathcal{X}^{\hat{=}} \land \varphi(\mathcal{X}), \quad \text{for any quantifier-free } \varphi(\mathcal{X}),\\
\tau(\exists\,\mathcal{Y}\,\varphi(\mathcal{X})) &= \exists\,\mathcal{Y}\,\tau(\varphi(\mathcal{X})),\\
\tau(\neg\,\varphi(\mathcal{X})) &= \mathcal{X}^{\hat{=}} \land \neg\,\tau(\varphi(\mathcal{X})) \ .
\end{aligned}$$

Observe that, as the translation stops after the prefix with quantifiers, there is no need to deal with binary connectives. For the quantitative claim, observe that $nk$ term in the time complexity $\mathcal{O}(nk+m)$ of the translation comes from the alternation of quantifiers. Every such alternation brings a linear size formula $\mathcal{Z}^{\hat{=}}$ where $\mathcal{Z}$ is the set of free variables in the subformula under the quantifier block. The qualitative claim is established as follows by structural induction.

**($\varphi(\mathcal{X})$ is quantifier-free)** Using Lemma 4.8, we have $\mathcal{G}^{\uparrow}(\tau(\varphi(\mathcal{X})) = \mathcal{G}^{\uparrow}(\mathcal{X}^{\triangleq} \wedge \varphi(\mathcal{X})) =$

$$= \mathcal{G}^{\uparrow}(\mathcal{X}^{\triangleq}) \cap \mathcal{G}^{\uparrow}(\varphi(\mathcal{X})) = \bigcup_{\substack{G\in\mathscr{L}(\mathcal{G})\\ G=(V,E,_,_)}} (V \cup E)^{\mathcal{X}} \cap \mathcal{G}^{\uparrow}(\varphi(\mathcal{X})) = \bigcup_{G\in\mathscr{L}(\mathcal{G})} G(\varphi(\mathcal{X})).$$

**($\exists\, \mathcal{X}\, \varphi(\mathcal{X})$)** We have We have $\mathcal{G}^{\uparrow}(\tau(\exists\, \mathcal{Y}\, \varphi(\mathcal{X})) = \mathcal{G}^{\uparrow}(\exists\, \mathcal{Y}\, \tau(\varphi(\mathcal{X}))) =$

$$= \pi_{\exists \mathcal{Y}}\big(\mathcal{G}^{\uparrow}(\tau(\varphi(\mathcal{X})))\big) \overset{\text{i.h.}}{=} \pi_{\exists \mathcal{Y}}\Big(\bigcup_{G\in\mathscr{L}(\mathcal{G})} G(\varphi(\mathcal{X}))\Big) = \bigcup_{G\in\mathscr{L}(\mathcal{G})} \pi_{\exists \mathcal{Y}}\big(G(\varphi(\mathcal{X}))\big) = \bigcup_{G\in\mathscr{L}(\mathcal{G})} G(\exists\, \mathcal{Y}\, \varphi(\mathcal{X})).$$

**($\neg\, \varphi(\mathcal{X})$)** Using Lemma 4.8 at ($\circledast$), we have $\mathcal{G}^{\uparrow}(\tau(\neg\, \varphi(\mathcal{X})) =$

$$\begin{aligned}
&= \mathcal{G}^{\uparrow}(\mathcal{X}^{\triangleq} \wedge \neg\, \tau(\varphi(\mathcal{X})))\\
&= \mathcal{G}^{\uparrow}(\mathcal{X}^{\triangleq}) \cap \mathcal{G}^{\uparrow}(\neg\, \tau(\varphi(\mathcal{X})))\\
&= \mathcal{G}^{\uparrow}(\mathcal{X}^{\triangleq}) \cap \big((\mathcal{V}(\mathcal{G}) \cup \mathcal{E}(\mathcal{G}))^{\mathcal{X}} \setminus \mathcal{G}^{\uparrow}(\tau(\varphi(\mathcal{X})))\big)\\
&= \big(\mathcal{G}^{\uparrow}(\mathcal{X}^{\triangleq}) \cap (\mathcal{V}(\mathcal{G}) \cup \mathcal{E}(\mathcal{G}))^{\mathcal{X}}\big) \setminus \big(\mathcal{G}^{\uparrow}(\mathcal{X}^{\triangleq}) \cap \mathcal{G}^{\uparrow}(\tau(\varphi(\mathcal{X})))\big)\\
&= \mathcal{G}^{\uparrow}(\mathcal{X}^{\triangleq}) \setminus \big(\mathcal{G}^{\uparrow}(\mathcal{X}^{\triangleq}) \cap \mathcal{G}^{\uparrow}(\tau(\varphi(\mathcal{X})))\big)\\
&= \mathcal{G}^{\uparrow}(\mathcal{X}^{\triangleq}) \setminus \mathcal{G}^{\uparrow}(\tau(\varphi(\mathcal{X})))\\
&\overset{\text{i.h.}}{=} \mathcal{G}^{\uparrow}(\mathcal{X}^{\triangleq}) \setminus \bigcup_{G\in\mathscr{L}(\mathcal{G})} G(\varphi(\mathcal{X}))\\
&\overset{(\circledast)}{=} \bigcup_{\substack{G\in\mathscr{L}(\mathcal{G})\\ G=(V,E,_,_)}} (V \cup E)^{\mathcal{X}} \setminus \bigcup_{G\in\mathscr{L}(\mathcal{G})} G(\varphi(\mathcal{X}))\\
&\overset{(*)}{=} \bigcup_{\substack{G\in\mathscr{L}(\mathcal{G})\\ G=(V,E,_,_)}} \big((V \cup E)^{\mathcal{X}} \setminus G(\varphi(\mathcal{X}))\big)\\
&= \bigcup_{G\in\mathscr{L}(\mathcal{G})} G(\neg\, \varphi(\mathcal{X}))\ .
\end{aligned}$$

Note that equality ($*$) is due to inclusions $G(\varphi(\mathcal{X})) \subseteq (V \cup E)^{\mathcal{X}}$, for every $G = (V, E, _, _)$ in $\mathscr{L}(\mathcal{G})$, whereas the uniqueness of derivation guarantees that that every pair of distinct graphs $G_1 \neq G_2$ of $\mathscr{L}(\mathcal{G})$ is disjoint, viz., $(V_1 \cup E_1) \cap (V_2 \cup E_2) = \varnothing$ for $G_1 = (V_1, E_1, _, _)$ and $G_2 = (V_2, E_2, _, _)$. ☐

**Corollary 6.4.** $\mathrm{FO}(\Gamma \cup \Sigma)$*-language checking problem is decidable D0L graph grammars even with non-terminals.*

*Proof.* Let $\varphi$ be a sentence in $\mathrm{FO}(\Gamma \cup \Sigma)$ and $\mathcal{G}$ be a D0L graph grammar possibly using non-terminals labels in $M$. Consider $\mathcal{G}^{¡!}$ obtained by completing $\mathcal{G}$ according to Prop. 4.2 and let $\tau_{¡!}$ be the translation of Prop. 6.3 for $N = \{¡, !\}$, namely $\tau_{¡!}(\varphi) \coloneqq \varphi \wedge \neg\, \exists\, x\, ¡(x) \wedge \neg\, \exists\, x y z\, !(x, y, z)$, where ¡ is a vertex label and ! is an edge label. Let $\tau_M$ be the translation of Prop. 6.3 for $N = M$. Let $\tau_0$ be the translation of Thm. 6.2. Then

there exists $G \in \mathscr{L}(\mathcal{G})_M$ such that $G \models \varphi$
iff there exists $G \in \mathscr{L}(\mathcal{G})$ such that $G \models \tau_M(\varphi)$ — by Prop. 6.3
iff there exists $G \in \mathscr{L}(\mathcal{G}^{¡!})_{\{¡,!\}}$ such that $G \models \tau_M(\varphi)$ — as $\mathscr{L}(\mathcal{G}) = \mathscr{L}(\mathcal{G}^{¡!})_{\{¡,!\}}$ by Prop. 4.2
iff there exists $G \in \mathscr{L}(\mathcal{G}^{¡!})$ such that $G \models \tau_{¡!}(\tau_M(\varphi))$ — by Prop. 6.3
iff $(\mathcal{G}^{¡!})^{\uparrow} \models \tau_0(\tau_{¡!}(\tau_M(\varphi)))$ — by Thm. 6.2.

☐

# D Automatic presentation of $\mathcal{G}^{\uparrow}$ in Section 7 is correct for complete grammars

This appendix brings together missing proofs of Sect. 7.

**Lemma 7.1.** *If $\mathcal{G}$ is complete, then* $[\![\mathscr{A}(\mathsf{vrt})]\!] = \mathcal{G}^{\uparrow}(\mathsf{vrt})$.

For facilitating the proof of this lemma, we need one additional lemma. We also recall the following notation: $w(\le i) = a_1 a_2 \ldots a_i$ when $w = a_1 a_2 \ldots a_n$ is a word over an alphabet, say $\Xi$, and $a_i \in \Xi$ for $i \in [n]$. Note that we omit subscript $\mathcal{G}$ in derivation related notations and that the dashed arrow represents the existance of a derivation of specified length.

**Lemma D.1.** *For all* $(V', E', _, _) \in \mathscr{L}_l(\mathcal{G})$ *and* $n \in [l-1]$

(1) *there exists a unique* $(V, E, _, _) \in \mathscr{L}_{l-n}(\mathcal{G})$ *such that* $\mathsf{A}_{\mathcal{G}} \xrightarrow{l-n} (V, E, _, _) \xrightarrow{n} (V', E', _, _)$,

(2) $u'(\le l-n) \in V \cup E$ *for every* $u' \in V' \cup E'$ *and* $u'(\le l-n) \xrightarrow{\mathrm{prd}(u'(l-n+1))\ldots\mathrm{prd}(u'(l))} u'$,

(3) *in particular,* $\varepsilon \xrightarrow{\mathrm{prd}(u'(1))\ldots\mathrm{prd}(u'(l))} u'$ *where* $\mathrm{prd}(u'(1)) = \mathsf{A}_{\mathcal{G}} \to _$,

(4) *for every* $u \in V \cup E$ *there exists* $v \in (\Theta \cup \Omega)^n$ *such that* $uv \in V' \cup E'$.

*Proof.* Follows directly from the definition of the derivation according to $\mathcal{G}$. □

With the help of the above lemma, the proof of Lemma 7.1 goes as follows.

*Proof of Lemma 7.1.* Let $\mathcal{G} \in 0\mathrm{L}(\Gamma, \Sigma, \Theta, \Omega)$ be a complete graph grammar. Because $\mathsf{ar}(\mathsf{vrt}) = 1$ and ⎕ is not used in $\Delta_{\mathsf{vrt}}$, one has $[\![\mathscr{A}(\mathsf{vrt})]\!] = \mathscr{L}(\mathscr{A}(\mathsf{vrt}))$.

(⊆) Let $v \in [\![\mathscr{A}(\mathsf{vrt})]\!]$. We show by induction on $l = |v|$ that there exists $(V_l, _, _, _) \in \mathscr{L}_l(\mathcal{G})$ such that $v \in V_l$. The inclusion follows then from $V_l \subseteq \mathcal{V}(\mathcal{G}) = \mathcal{G}^{\uparrow}(\mathsf{vrt})$.

- **($l = 1$)** Then $\iota_{\mathsf{vrt}} \xrightarrow{v} v \in \Delta_{\mathsf{vrt}}$ for some $\mathsf{A}_{\mathcal{G}} \to (V_1, _, _, _) \in \mathsf{V}_{\mathcal{G}}$ such that $v \in V_1$.
- **($l > 1$)** Let $uij = v$ for some $i, j \in \Theta$. By construction of $\mathscr{A}(\mathsf{vrt})$, there is a path $\iota_{\mathsf{vrt}} \overset{ui}{\dashrightarrow} i$ in $\mathscr{A}(\mathsf{vrt})$. Then $ui \in \mathscr{L}(\mathscr{A}(\mathsf{vrt}))$ because $i \in F_{\mathsf{vrt}}$. By induction hypothesis there exists $G_{l-1} = (V_{l-1}, E_{l-1}, \ell_{l-1}, \daleth_{l-1}) \in \mathscr{L}_{l-1}(\mathcal{G})$ such that $ui \in V_{l-1}$. As $uij \in \mathscr{L}(\mathscr{A}(\mathsf{vrt}))$, there is a transition $i \xrightarrow{j} j \in \Delta_{\mathsf{vrt}}$ such that $\mathrm{prd}(j) = P = \ell(i) \to _$. Then, one can consider label preserving map $h\colon V_{l-1} \to \mathsf{V}_{\mathcal{G}}$ such that $h(v) = P$. According to Lemma 4.1, this map extends into homomorphism $h\colon G_{l-1} \to \mathcal{G}$. The latter produces from $G_{l-1}$ a graph in $\mathscr{L}_l(\mathcal{G})$, say $(V_l, _, _, _)$, such that $v = uij \in V_l$.

(⊇) Let $v \in \mathcal{V}_l(\mathcal{G}) \subseteq \mathcal{G}^{\uparrow}(\mathsf{vrt})$. Thus, there exists $(V_l, _, _, _) \in \mathscr{L}_l(\mathcal{G})$ such that $v \in V_l$. By (2) and (3) of Lemma D.1, we have $\varepsilon \xrightarrow{\mathrm{prd}(v(1))\ldots\mathrm{prd}(v(l))} v$ where $\mathrm{prd}(v(1)) = \mathsf{A}_{\mathcal{G}} \to _$ and, for every $k \in [l-1]$, we have $v(\le k) \xrightarrow{\mathrm{prd}(v(k+1))} v(\le k+1)$ where $\mathrm{prd}(v(k+1)) = \ell(v(k)) \to _$. Then, the construction of $\mathscr{A}(\mathsf{vrt})$ allows us to affirm that $\iota_{\mathsf{vrt}} \xrightarrow{v(1)} \cdots \xrightarrow{v(l)} v(l)$ is an accepting path of $\mathscr{A}(\mathsf{vrt})$. Hence, $v \in \mathscr{L}(\mathscr{A}(\mathsf{vrt}))$. □

**Lemma 7.3.** *If* $\mathcal{G}$ *is complete, then* $(\!|\mathscr{A}_{\mathfrak{I}}|\!) = \mathfrak{I}(\mathcal{G})$*, where* $\mathfrak{I}(\mathcal{G})$ *is the set of incidences of* $\mathcal{G}^{\uparrow}$.

Remember that $\mathfrak{I}(\mathcal{G})$ is defined in Sect. 4 as $\mathfrak{I}(\mathcal{G}) \coloneqq \bigcup_{(_,_,_,\daleth) \in \mathscr{L}(\mathcal{G})} \daleth$. Notation $(\!|\mathscr{A}_{\mathfrak{I}}|\!)$ is introduced in Sect. 7 just above Lemma 7.3 together with a view of $\overrightarrow{\Sigma} \cup \overleftarrow{\Sigma}$ as a family of mappings. To keep the proof simple, all orientations of $\overleftrightarrow{a}$ and $\overleftrightarrow{b}$ are taken as in the construction of $\mathscr{A}_{\mathfrak{I}}$ without being explicitly mentioned. If instead orientations were set explicitly, there would be four similar cases to consider in the proof.

*Proof.* Let $\mathcal{G} \in 0\mathrm{L}(\Gamma, \Sigma, \Theta, \Omega)$ be a complete graph grammar. Observe that ⎕ is not used in $\Delta_{\mathfrak{I}}$. Consequently every path of length $n$ of $\mathscr{A}_{\mathfrak{I}}$ is labelled by some $\otimes(s_1, s_2, s_3)$ such that $s_1, s_2, s_3 \in (\Theta \cup \Omega)^n$.

(⊆) Let $\overleftrightarrow{a}(u, w, v) \in (\!|\mathscr{A}_{\mathfrak{I}}|\!)$. We show by induction on $l = |u| = |v| = |w|$ that there exists $(_, _, _, \daleth_l) \in \mathscr{L}_l(\mathcal{G})$ such that $\overleftrightarrow{a}(u, w, v) \in \daleth_l$. The inclusion follows then from $\daleth_l \subseteq \mathfrak{I}(\mathcal{G})$.

- **($l = 1$)** Then $\iota_{\mathfrak{I}} \xrightarrow{(u,w,v)} (\overrightarrow{a}, u, w, v) \in \Delta_{\mathfrak{I}}$ for some $\mathsf{A}_{\mathcal{G}} \to (_, _, _, \daleth_1) \in \mathsf{V}_{\mathcal{G}}$ such that $a(u, w, v) \in \daleth_1$.

**$(l>1)$** Let $(u'i'i, w'j'j, v'k'k) = (u,w,v)$ for some $i,i',k,k' \in \Theta$, $j \in \Omega$ and $j' \in \Theta \cup \Omega$. By construction of $\mathscr{A}_\beth$, there is a path $\iota_\beth \overset{\otimes(u'i',w'j',v'k')}{\dashrightarrow} (\beta, i', q, k')$ in $\mathscr{A}_\beth$ where $q = \square$ and $\beta = \diamond$ or, $q = j'$ and $\beta = \overleftrightarrow{b}$, viz., $\beta \in \{\overrightarrow{b}, \overleftarrow{b}\}$, for some $b \in \Sigma$.

- If $q = \square$ then the above path can be more precisely written as
$$\iota_\beth \xrightarrow{(r_1,r_1,r_1)} (\diamond, r_1, \square, r_1) \xrightarrow{(r_2,r_2,r_2)} \cdots \xrightarrow{(r_{l-1},r_{l-1},r_{l-1})} (\diamond, r_{l-1}, \square, r_{l-1}) \qquad (*)$$
where $i' = j' = k' = r_{l-1}$ and $u' = v' = w' = r_1 \dots r_{l-2}$ with $\{r_1, \dots, r_{l-1}\} \subseteq \Theta$. This path mimics path $\iota_{\mathsf{vrt}} \xrightarrow{r_1} r_1 \xrightarrow{r_2} \cdots \xrightarrow{r_{l-1}} r_{l-1}$ of $\mathscr{A}(\mathsf{vrt})$. By Lemma 7.1, this is an accepting path. Thus $u'i' = w'j' = v'k' = \overline{r}$, where $\overline{r} := r_1 \dots r_{l-1}$, is in $\mathcal{V}_{l-1}(\mathcal{G})$. As $\mathscr{A}_\beth$ is deterministic and $\overleftrightarrow{a}(u,w,v) \in (\!|\mathscr{A}_\beth|\!)$, path $(*)$ extends in $\mathscr{A}_\beth$ into the path accepting $\otimes(u,w,v)$. Consequently, there is $(\diamond, r_{l-1}, \square, r_{l-1}) \xrightarrow{(i,j,k)} (\overrightarrow{a}, i, j, k) \in \Delta_\beth$ such that $\ell(r_{l-1}) \to (_,_,_,\beth) \in \mathsf{V}_\mathcal{G}$ and $a(i,j,k) \in \beth$. By definition of the derivation, there exists $G_l = (V_l, E_l, \ell_l, \beth_l)$ such that $a(\overline{r}i, \overline{r}j, \overline{r}k) \in \beth_l$. Hence, $a(u,w,v) = (\overline{r}i, \overline{r}j, \overline{r}k) \in \mathfrak{I}_l(\mathcal{G})$.
- If $q = j'$, then $\iota_\beth \overset{\otimes(u'i',w'j',v'k')}{\dashrightarrow} (\overleftrightarrow{b}, i', j', k')$ is an accepting path of $\mathscr{A}_\beth$, viz., $\overleftrightarrow{b}(u'i', w'j', v'k') \in (\!|\mathscr{A}_\beth|\!)$. By induction hypothesis there exists $G_{l-1} = (_,_,\ell_{l-1},\beth_{l-1}) \in \mathscr{L}_{l-1}(\mathcal{G})$ such that $\overleftrightarrow{b}(v'k', w'j', u'i') \in \beth_{l-1}$. As $\otimes(u,w,v) \in \mathscr{L}(\mathscr{A}_\beth)$, there is a transition $(\overleftrightarrow{b}, i', j', k') \xrightarrow{(i,j,k)} (\overleftrightarrow{a}, i, j, k) \in \Delta_\beth$. By construction of $\mathscr{A}_\beth$, there exist $\mathrm{prd}(i) = \ell_{l-1}(i') \to _, \mathrm{prd}(k) = \ell_{l-1}(k') \to _ \in \mathsf{V}_\mathcal{G}$ and $\mathrm{prd}(j) = \mathrm{prd}(i) \xrightarrow{\overleftrightarrow{b}} \mathrm{prd}(k) \to \beth \in \mathsf{E}_\mathcal{G}$ such that $\overleftrightarrow{a}(i,j,k) \in \beth$. Then, one can consider label preserving map $h\colon V_{l-1} \to \mathsf{V}_\mathcal{G}$ such that $h(u'i') = \mathrm{prd}(i)$, $h(v'k') = \mathrm{prd}(k)$. Observe that this homomorphism can be chosen so that $h(w'j') = \mathrm{prd}(j)$. Consequently, $h$ produces from $G_{l-1}$ a graph in $\mathscr{L}_l(\mathcal{G})$, say $(_,_,_,\beth_l)$, such that $\overleftrightarrow{a}(u,w,v) \in \beth_l$.

$(\supseteq)$ Let $\overleftrightarrow{a}(u,w,v) \in \mathfrak{I}_l(\mathcal{G})$. Thus, there exists $(_,_,_,\beth_l) \in \mathscr{L}_l(\mathcal{G})$ such that $\overleftrightarrow{a}(u,w,v) \in \beth_l$. We show by induction on $l \in \mathbb{N}_{>0}$ that $\iota_\beth \overset{\otimes(u,w,v)}{\dashrightarrow} (\overleftrightarrow{a}, _,_,_)$ is a path in $\mathscr{A}_\beth$. Note that such a path is accepting.

**$(l=1)$** As $\overleftrightarrow{a}(u,w,v) \in \beth_1$ for some $G_1 = (_,_,_,\beth_1) \in \mathscr{L}_1(\mathcal{G})$, there is production $\mathsf{A}_\mathcal{G} \to G_1 \in \mathsf{V}_\mathcal{G}$. Then more precisely, $\overrightarrow{a}(u,w,v) = a(u,w,v) \in \beth_1$. Consequently, $\iota_\beth \xrightarrow{(u,w,v)} (\overrightarrow{a}, u, w, v)) \in \Delta_\beth$.

**$(l>1)$** As $\overleftrightarrow{a}(u,w,v) \in \beth_l$ for some $G_l = (_, E_l, _, \beth_l) \in \mathscr{L}_l(\mathcal{G})$, by Lemma D.1(1), there exists $G_{l-1} = (V_{l-1}, _, \ell_{l-1}, \beth_{l-1}) \in \mathscr{L}_{l-1}(\mathcal{G})$ such that $G_{l-1} \xrightarrow{h} G_l$ for some homomorphism $h\colon G_{l-1} \to \mathcal{G}$. As $w \in E_l$ is either innate or inherited, we have two cases.

- $(u,w,v) = (si, sj, sk)$ for some $i,k \in V_0$, $j \in E_0$ and $s \in V_{l-1}$ such that $G_0 = (V_0, E_0, _, \beth_0)$ is the right hand side of vertex production $h(s) = \ell_{l-1}(s) \to G_0$. Moreover $a(i,j,k) \in \beth_0$. By Lemma 7.1, $\iota_{\mathsf{vrt}} \overset{s}{\dashrightarrow} i' = s(l-1)$ is an accepting path in $\mathscr{A}(\mathsf{vrt})$. It is mimicked in $\mathscr{A}_\beth$ by $\iota_\beth \overset{\otimes(s,s,s)}{\dashrightarrow} (\diamond, i', \square, i')$. From there one more step can be done as, by construction, $\{(\diamond, i', \square, i') \xrightarrow{(i,j,k)} (\overrightarrow{a}, i, j, k) \in \Delta_\beth$. Consequently, $\overrightarrow{a}(u,w,v) = \overrightarrow{a}(si,sj,sk) \in (\!|\mathscr{A}_\beth|\!)$.
- $(u,w,v) = (u'i, w'j, v'k)$ where $\overleftrightarrow{b}(u',w',v') \in \beth_{l-1}$ and $\overleftrightarrow{a}(i,j,k) \in \beth$ for some edge production $h(w') = \mathrm{prd}(i) \xrightarrow{\overleftrightarrow{b}} \mathrm{prd}(k) \to \beth$. By induction hypothesis, $\overleftrightarrow{b}(u',w',v') \in (\!|\mathscr{A}_\beth|\!)$. Consequently, $\iota_\beth \underset{\mathscr{A}_\beth}{\overset{\otimes(u',w',v')}{\dashrightarrow}} (\overleftrightarrow{b}, i', j', k')$ where $i' = u'(l-1)$, $j' = w'(l-1)$ and $k' = v'(l-1)$. By construction, $(\overleftrightarrow{b}, i', j', k') \xrightarrow{(i,j,k)} (\overleftrightarrow{a}, i, j, k) \in \Delta_\beth$. Therefore, the above path extends into $\iota_\beth \underset{\mathscr{A}_\beth}{\overset{\otimes(u'i,w'j,v'k)}{\dashrightarrow}} (\overleftrightarrow{a}, i, j, k)$. Hence, $\overrightarrow{a}(u,w,v) = \overrightarrow{a}(u'i, w'j, v'k) \in (\!|\mathscr{A}_\beth|\!)$.

$\square$

**Lemma 7.6.** *If $\mathcal{G}$ is complete, then* $[\![\mathscr{A}(\between_0)]\!] = \between_0$.

*Proof.* Since $\mathscr{B}_1$ accepts all elements of $\mathcal{V}(\mathcal{G}) \cup \mathcal{E}(\mathcal{G})$ duplicated, after switching via $\Delta$ into $\mathscr{B}_2$ every pair $(w, w) \in [\![\mathscr{B}_1]\!]$ extends into pair $(wu, wv)$ such that $(u, v) \in [\![\mathscr{B}_2]\!]$, where $(u, v)$ is any pair in $(\mathcal{V}_l(\mathcal{G}) \cup \mathcal{E}_l(\mathcal{G}))^2$, for $l \in \mathbb{N}$. However, the switch is activated by any pair of distinct productions sharing the same left hand side. This behaviour matches precisely the definition of $\between_0$. $\square$